\documentclass[12pt]{article}
\usepackage{amssymb,bbold}
\usepackage{epsfig}
\usepackage{bm}

\newcommand{\sect}[1]{\setcounter{equation}{0}\section{#1}}
\renewcommand{\theequation}{\arabic{section}.\arabic{equation}}
\newcommand{\s}{\sigma}
\newcommand{\e}{\epsilon}
\newcommand{\bea}{\begin{eqnarray}}
\newcommand{\eea}{\end{eqnarray}}
\newcommand{\nn}{\nonumber \\}
\newcommand{\p}[1]{(\ref{#1})}

\newcommand{\balpha}{{\mbox{\boldmath $\alpha$}}}

\newcommand{\bgamma}{{\mbox{\boldmath $\gamma$}}}

\newcommand{\blambda}{{\mbox{\boldmath $\lambda$}}}

\newcommand{\bchi}{{\mbox{\boldmath $\chi$}}}
\newcommand{\bomega}{{\mbox{\boldmath $\omega$}}}

\newcommand{\cross}{\times}

\def\be{\begin{equation}}
\def\ee{\end{equation}}
\def\ba{\begin{eqnarray}}
\def\ea{\end{eqnarray}}
\def\ff{{\cal{F}}}
\def\godpar{\gamma}

\font\mybb=msbm10 at 11pt 
\def\bb#1{\hbox{\mybb#1}}

\def\bR {\bb{R}}
\def\bE {\bb{E}}

\def\bH {\bb{H}}
\def\bC {\bb{C}}

\def\cL{{\cal L}}

\def\pd{\partial_}

\def\pd{\partial_}
\def\e{\epsilon}
\def\g{\gamma}

\begin{document}

\baselineskip 18pt

\begin{titlepage}

\vfill
\begin{flushright}
\today\\
QMUL-PH-02-13\\
hep-th/0209114\\
\end{flushright}

\vfill

\begin{center}
{\bf \Large All supersymmetric solutions of minimal supergravity}\\
\vspace*{3mm} {\bf \Large in five dimensions}

\vskip 10.mm {Jerome P. Gauntlett$^{1}$, Jan B.
Gutowski$^{2}$,
Christopher M. Hull$^{3}$,}\\
{Stathis Pakis$^{4}$
and Harvey S. Reall$^{5}$}\\
\vskip 1cm

{\it
Department of Physics\\
Queen Mary, University of London\\
Mile End Rd, London E1 4NS, UK
}\\

\vspace{6pt}

\end{center}
\par

\begin{abstract}
\noindent All purely bosonic supersymmetric solutions of
minimal supergravity in five dimensions are classified.
The solutions preserve either one half or all of the supersymmetry.
Explicit examples of new solutions are
given, including a  large family of plane-fronted waves
and a maximally supersymmetric analogue of the G\"odel universe which
lifts to a solution of eleven dimensional supergravity that preserves
$20$ supersymmetries.

\end{abstract}
\vskip 1cm
\centerline{\it Dedicated to the memory of Sonia Stanciu}
\vfill \vskip 5mm \hrule width 5.cm \vskip 5mm {\small
\noindent $^1$ E-mail: j.p.gauntlett@qmul.ac.uk \\
\noindent $^2$ E-mail: j.b.gutowski@qmul.ac.uk \\
\noindent $^3$ E-mail: c.m.hull@qmul.ac.uk \\
\noindent $^4$ E-mail: s.pakis@qmul.ac.uk\\
\noindent $^5$ E-mail: h.s.reall@qmul.ac.uk }
\end{titlepage}

\sect{Introduction}

Solutions of higher dimensional supergravity theories have played a
key role in elucidating the structure of string theory.
Many interesting solutions have been found describing
higher dimensional black holes, black branes and
their intersections, pp-waves and so on. However, two
recent discoveries suggest that higher dimensional gravity may harbour
a much richer spectrum of objects that remains to be discovered.

First, it has been suggested that there may exist
a family of black brane solutions without translational
symmetry \cite{Horowitz:2001cz,Horowitz:2002ym}.
There are no known exact solutions describing such objects
but there is numerical evidence \cite{Gubser:2001ac,wiseman}
that such solutions do
exist. Secondly, an exact solution of the five dimensional vacuum
Einstein equations has been found that describes an asymptotically flat
black hole of topology $S^1 \times S^2$: a rotating black ring
\cite{Emparan:2001wn}. This
is the first example of a black hole of non-spherical topology.
Furthermore, the existence of this solution implies that the
black hole uniqueness theorems cannot be
extended to five dimensions, except in the static case
\cite{Gibbons:2002av}.

It is tempting to suspect that these new solutions are just the tip of
the iceberg, and that many more surprises will be found in higher
dimensions. It is clearly desirable to have a better understanding of
exact solutions of higher dimensional supergravity
theories. Unfortunately, solving the Einstein equations is notoriously
difficult even in four dimensions.

Supersymmetric solutions of supergravity theories are of
particular importance in string theory because such solutions often
have certain stability and non-renormalization properties that are
not possessed by non-supersymmetric solutions. For example,
it has been possible to
give a microscopic description of certain supersymmetric
black holes \cite{entropy}.
However, this work relies on the assumption that there is
a uniqueness theorem for supersymmetric black holes in four and five
dimensions. For supersymmetric rotating black holes (which only seem
to exist in five dimensions), this might
not be the case if supersymmetric black rings
exist. It is therefore of interest to examine the extent to which
supersymmetry excludes some of the more exotic solutions of higher
dimensional gravity discussed above, or alternatively provides a setting
in which they can be studied in more detail. To do so, we would like to
know
the general nature of supersymmetric solutions of higher dimensional
supergravity theories.

Although there are many partial results for $D=10$ and $D=11$
supergravities concerning manifolds with special holonomy, various brane
solutions etc, a systematic classification of {\it all}
supersymmetric solutions remains a challenging problem.
A more modest goal is to attempt a similar classification
for simpler supergravity theories, which can be viewed as a truncation
of the $D=10$ or $D=11$ supergravity theories.

Some time ago, following \cite{GibbonsHull}, this was carried out for
minimal $N=2$ supergravity in $D=4$ by Tod \cite{Tod}. It was shown
that the
supersymmetric solutions fall into two classes depending on
whether the Killing vector obtained from the Killing
spinor is time-like or null.
Moreover, the general solution could be obtained explicitly in
both cases. In the timelike case, one obtains the
Israel-Wilson-Perjes (IWP) class of solutions and the null case
consists of pp-waves.
There are also supersymmetric solutions with sources, provided the
sources
saturate a BPS bound relating the energy density to the electric and
magnetic charge densities  \cite{GibbonsHull}, \cite{Tod}.
Some generalizations of this result for other $D=4$
theories were presented in \cite{tod:95}.

The goal of the present paper is to extend this classification to the
simplest higher dimensional supergravity theory: the minimal $N=1$,
$D=5$ supergravity theory constructed in \cite{cremmer:81}. This
is a similar theory in the sense that it has the same number of
supercharges and furthermore after dimensional reduction on a
circle it gives $N=2$ supergravity in $D=4$ coupled to a vector
multiplet. However, in five dimensions it is not possible to use the
Newman-Penrose formalism adopted in \cite{Tod,tod:95} and
new techniques are required.

Following \cite{Gauntlett:2001ur,Gauntlett:2002sc},
the basic strategy is to assume the
existence of at least one Killing spinor, and consider the
differential forms that can be constructed as bilinear quantities
from this spinor. These satisfy a number of algebraic and
differential conditions that then can be used to deduce the form
of the metric and the gauge fields.
It is clear from the outset that for this theory one should not
expect to be able to explicitly construct all
solutions in closed form, since,  for example, a simple class of
solutions is the product of $K3$ with a flat time direction with
vanishing gauge fields, and the explicit metric on $K3$ is not known.
Nevertheless, we are able to give a simple set of rules for the
construction of all supersymmetric solutions in this theory.

We find that the supersymmetric solutions fall into two
classes, as in \cite{Tod}, depending on whether the Killing
vector $\bar{\epsilon} \gamma^{\mu} \epsilon$ obtained from the
Killing spinor $\epsilon$ is timelike or null. In each class the
solutions
preserve $1/2$ or all of the supersymmetry. In the null case, the
general
solution can be obtained explicitly.
It is a plane-fronted wave, specified by three
arbitrary harmonic functions on $\bR^3$. This can be contrasted with
the situation in the $N=2$, $D=4$ theory, where the null solutions
are given by pp-waves. In our solution, pp-waves appear merely as
a special case specified by two harmonic functions on $\bR^3$.
Perhaps somewhat surprisingly, our null family of solutions
contains some familiar {\it static} spacetimes such as
the supersymmetric magnetic black string solution
\cite{Gibbons:1994vm}, and its near horizon geometry, $AdS_3 \times
S^2$. The point is that these spacetimes are boost invariant and
therefore admit a null Killing vector field, and it turns out that
this is what is obtained from a Killing spinor.

In the timelike case we
find that supersymmetric solutions are specified by the following
data: a hyper-K\"ahler $4$-manifold $B$ describing the spatial
base geometry orthogonal to the orbits of the Killing vector field;
a $1$-form connection $\omega$ defined locally on $B$ and a
function $f$ defined globally on $B$, satisfying a pair of simple
equations. Solutions with non-vanishing $\omega$
generically describe rotating, or boosted, spacetimes.

The electrically charged rotating supersymmetric black hole of
Beckenridge, Myers, Peet and Vafa (BMPV) \cite{bmpv} can be
obtained as a solution of $N=1$, $D=5$ supergravity
\cite{gauntlett:99}. In our classification, it has base manifold
$B=\bR^4$.
This solution has some rather peculiar properties.
{}For example, although the solution has angular momentum, the
angular velocity of the horizon is zero \cite{gauntlett:99}.
{}Furthermore, there are closed timelike curves (CTCs) behind the
horizon \cite{bmpv}, and if the angular momentum is sufficiently
large then the solution no longer describes a black hole but is
instead a geodesically complete asymptotically flat,
supersymmetric time machine \cite{Gibbons:1999uv}. The appearance of
naked CTCs was related to a breakdown in unitarity of the
underlying microscopic description in \cite{herdeiro:00}.

One result of our investigation is that CTCs seem to be generic
for the timelike solutions in five dimensions (this is similar to $D=4$
as
closed time-like curves are generic for the $D=4$ IWP solutions
\cite{hartle:72}). It turns out to be rather difficult to find any
new solutions which do {\it not} have closed timelike curves or
singularities. One of our solutions is of particular interest
owing to its close similarity to G\"odel's four dimensional
rotating universe solution \cite{godel}. G\"odel's solution
motivated interest in time machines in General Relativity because
it is a homogeneous solution with trivial topology $\bR^4$ yet
contains CTCs through every point.  Our five dimensional solution
has very similar properties and is slightly simpler than
G\"odel's.

It is interesting to note that there are some
similarities with the equations arising in the time-like case and
those in the ``Resolution through transgression'' series of papers
(see \cite{cveticetal} for a review). In particular, if the
solutions are static then a harmonic function on
the hyper-K\"ahler base appears in the $D=5$ solution.
Generically these are singular solutions in $D=5$. Stationary
solutions, on the other hand, have the harmonic function replaced by a
function that solves a Laplace equation modified by the square of a
self-dual
harmonic two form, and we show that these can lead to non-singular
solutions.

{}For the timelike case with base space given by a Gibbons-Hawking
space (the most general with a tri-holomorphic Killing vector),
we are able to show that the most general solution is specified
by four arbitrary harmonic functions on $\bR^3$.

We have examined the further conditions required for our solutions to
preserve maximal supersymmetry.
For the null case we are led to flat space,
$AdS_3\times S^2$ and a certain plane-wave solution
\cite{Meesn}. In the timelike case, flat space, $AdS_2\times
S^3$ and the near horizon geometry of the BMPV solution
are all known to be maximally supersymmetric but surprisingly it turns
out that the generalized G\"odel solution also preserves all eight
supercharges.

The maximally supersymmetric timelike solutions just
listed all have flat base space. However, maximally
supersymmetric solutions can also be obtained from non-flat base
spaces. For example, there is a novel construction of $AdS_2\times S^3$
using
a nakedly singular Eguchi-Hanson space and another solution using
negative mass
Taub-NUT that gives rise to the G\"odel spacetime. These examples show
that the five dimensional geometry does not uniquely determine the
base space of maximally supersymmetric solutions. Furthermore, it
turns out that the maximally supersymmetric null solutions have some
Killing spinors that correspond to timelike Killing vectors and hence
these solutions must lie in {\it both} classes.
In the timelike description, the plane wave arises from a base
space describing a smeared distribution of Taub-NUT instantons and
$AdS_3 \times S^2$ arises from another singular hyper-K\"ahler base
space.
These examples illustrate that physically interesting and regular
five dimensional solutions can arise from a pathological base space.

All of the solutions of minimal $D=5$ supergravity can be uplifted
to obtain solutions of $D=10$ and $D=11$ supergravity. In general one
expects that
the uplifted solutions will preserve either 4 or 8 supersymmetries for
the
generic and the maximally
supersymmetric D=5 solutions, respectively. Surprisingly, we show that
the
G\"odel solution uplifts to a solution of $D=11$ supergravity that
preserves $20$ supersymmetries. Although general arguments have been
put forward
for the existence of supergravity solutions preserving all fractions
of supersymmetry, and in particular between $1/2$ and $1$
\cite{Gauntlett:1999dt},
to date the only such solutions that have been found are in the plane
wave category
\cite{Cvetic:2002hi}-\cite{Michelson:2002ps}.
This G\"odel solution thus constitutes a new class of solution
preserving an exotic fraction of supersymmetry.

Our results constitute the first analysis of {\it all} supersymmetric
solutions of a higher dimensional supergravity theory. The results of
this work  provide encouraging evidence that the strategy of
\cite{Gauntlett:2001ur,Gauntlett:2002sc} could be used to perform a
similar classification of all supersymmetric solutions of other higher
dimensional supergravity theories. A key idea of
\cite{Gauntlett:2001ur,Gauntlett:2002sc}
is to relate supersymmetric solutions to different kinds of
$G$-structures and we will discuss this relationship for $D=5$.
One motivation for studying the minimal $D=5$ supergravity is that it
is very similar in structure to 11-dimensional supergravity.
Supersymmetric solutions to D=5, N=1 supergravity coupled to matter
have been studied in e.g. \cite{chamsab:98}, \cite{sabr:98}.

The plan of the rest of the paper is as follows. In section
\ref{sec:basics} we show how various bosonic quantities are
constructed from a Killing spinor and derive differential and
algebraic relations between these quantities. Section
\ref{sec:timelike} analyses
the case when the Killing vector constructed from the Killing spinor
is timelike and includes several new solutions.
Section \ref{sec:null} carries out a similar analysis when the Killing
vector
is null. Section \ref{sec:maximal} discusses maximally
supersymmetric solutions. Section 6
discusses the connection with $G$-structures.
Section \ref{sec:godel11} uplifts the G\"odel solution to D=11
supergravity
and section \ref{sec:discussion} concludes.
The paper contains two appendices.

\sect{D=5 supergravity}

\label{sec:basics}

The bosonic action for minimal supergravity in five dimensions is
\be \label{eqn:action}
  S = \frac{1}{4\pi G} \int \left( -\frac{1}{4} R * 1 - \frac{1}{2} {}F
  \wedge *{}F - \frac{2}{3\sqrt{3}} {}F \wedge {}F \wedge A \right),
\ee We will adopt the conventions of \cite{cremmer:81} and these
are outlined in appendix A\footnote{The sign of the Chern-Simons
term corrects that appearing in \cite{cremmer:81}.}.
The bosonic equations of motion are
\bea\label{eqofmot}
R_{\alpha\beta}+2({}F_{\alpha\gamma}{}F_\beta{}^\gamma
-\frac{1}{6}g_{\alpha\beta}{}F^2)&=&0\nn d*{}F +
\frac{2}{\sqrt{3}} {}F \wedge {}F&=0&
\eea
where ${}F^2\equiv {}F_{\alpha\beta}{}F^{\alpha\beta}$. A bosonic
solution to the equations of motion is supersymmetric if it admits
a a super-covariantly constant spinor obeying
\be\label{kspin}
  \left[ D_\alpha + \frac{1}{4\sqrt{3}}  \left(
  \gamma_\alpha{}^{\beta\gamma} - 4
\delta^{\beta}_{\alpha}\gamma^{\gamma}
   \right){}F_{\beta \gamma} \right] \epsilon^a = 0.
\ee where $\epsilon^a$ is a symplectic Majorana spinor. We shall call
such
spinors Killing spinors. Our
strategy for determining the most general bosonic supersymmetric
solutions\footnote{Note that there are spacetimes admitting a Killing
spinor
that do not satisfy the equations of motion.
These can be viewed as solutions of the field equations with
additional sources, and
supersymmetry imposes conditions on these sources.
For example, in the case of solutions with a timelike Killing vector,
the
source must be a charged dust with charge density equal to the mass
density,
analogous to the sources in \cite{GibbonsHull,Tod}. Here we will
restrict
ourselves to solutions of the field equations without sources.}
is to analyse the differential forms that can be
constructed from commuting Killing spinors. We first investigate
algebraic
properties of these forms, and then their differential properties.

{}From a single commuting spinor $\epsilon^a$ we can construct a scalar
$f$,
a 1-form $V$ and three 2-forms $\Phi^{ab} \equiv \Phi^{(ab)}$: \be
  f \epsilon^{ab} = \bar{\epsilon}^a \epsilon^b,
\ee \be
  V_\alpha \epsilon^{ab} = \bar{\epsilon}^a \gamma_\alpha \epsilon^b,
\ee \be
  \Phi^{ab}_{\alpha \beta} = \bar{\epsilon}^a \gamma_{\alpha \beta}
  \epsilon^b,
\ee $f$ and $V$ are real, but $\Phi^{11}$ and $\Phi^{22}$ are
complex conjugate and $\Phi^{12}$ is imaginary. It is   useful
to work with three real two-forms defined by \be
  \Phi^{(11)} = X^{(1)} + i X^{(2)}, \qquad
  \Phi^{(22)} = X^{(1)} - i X^{(2)}, \qquad
  \Phi^{(12)} = - i X^{(3)}.
\ee These quantities give a total of $1+5+3\times 10 = 36$ real
degrees of freedom. To understand this, note that $\epsilon^a$ has
a total of $8$ real components. The product $\epsilon^a_{\alpha}
\epsilon^b_{\beta}$ is symmetric in $(a, \alpha)$ and $(b,
\beta)$. A symmetric $8 \times 8$ matrix has $36$ components,
corresponding to the $36$ degrees of freedom obtained above (by
projecting with $C$, $C\gamma_\alpha$ and $C
\gamma_{\alpha\beta}$) . Having said this, $\epsilon^a_{\alpha}
\epsilon^b_{\beta}$ is {\it not} a general symmetric $8 \times 8$
matrix because it has rank $1$ and only $8$ real degrees of
freedom. It follows that there should be algebraic relations
between the above quantities that reduce the number of independent
components from $36$ to $8$.

It will be useful to record some of these identities which can be
obtained from various {}Fierz identities. We first note that \be
  V_{\alpha} V^{\alpha} = f^2
\ee which implies that $V$ is timelike, null or zero. The final
possibility can be eliminated by noting $2 V_0 =
{\epsilon_a}^\dagger \epsilon_a > 0$ in any region in which the Killing
spinor
is non-vanishing.
We will work in such a region, and analytically continue the metric to
the fixed point sets of the
  isometry generated by $V$ at which $V$ vanishes.
In later sections we will analyse the time-like
and null cases separately. We also have \be \label{eqn:XwedgeX}
  X^{(i)} \wedge X^{(j)} = -2\delta_{ij} f * V,
\ee \be \label{eqn:VdotX}
  i_V X^{(i)} = 0,
\ee \be \label{eqn:VstarX}
  i_V * X^{(i)} = - f X^{(i)},
\ee \be \label{eqn:XcontX}
  X^{(i)}_{\gamma \alpha} X^{(j) \gamma}{}_{\beta} = \delta_{ij} \left(
  f^2 \eta_{\alpha\beta} - V_{\alpha} V_{\beta} \right) + \epsilon_{ijk}
f
  X^{(k)}_{\alpha\beta} \ ,
\ee where $\epsilon_{123} = +1$ and, for a vector $Y$ and $p$-form $A$,
$(i_Y A)_{\alpha_1
\ldots \alpha_{p-1}} \equiv Y^{\beta} A_{\beta \alpha_1 \ldots
\alpha_{p-1}}$. {}Finally, it is useful to record
\be \label{eqn:Vproj}
  V_{\alpha} \gamma^\alpha \epsilon^a = f \epsilon^a \ ,
\ee and \be \label{eqn:Phiproj}
  \Phi^{ab}_{\alpha \beta} \gamma^{\alpha \beta} \epsilon^c = 8f
\epsilon^{c(a} \epsilon^{b)}. \ee
{}For the  remainder of the paper we will usually suppress
the symplectic indices on the spinors.

We now turn to the differential conditions that can be obtained by
assuming that $\epsilon$ is a Killing spinor. We differentiate
$f$, $V$, $\Phi$ in turn and use \p{kspin}. Starting with $f$ we
find \be \label{eqn:df}
  df = -\frac{2}{\sqrt{3}} i_V {}F.
\ee
Taking the exterior derivative and using the
Bianchi identity for ${}F$ then gives \be \label{eqn:lie{}F}
  {\cal L}_V {}F = 0,
\ee where ${\cal L}$ denotes the Lie derivative. Next,
differentiating $V$ gives \be\label{deevee}
  D_\alpha V_\beta = \frac{2}{\sqrt{3}} {}F_{\alpha \beta} f + \frac{1}{2
  \sqrt{3}} \epsilon_{\alpha \beta \gamma \delta \epsilon} {}F^{\gamma
  \delta} V^{\epsilon},
\ee which implies $ D_{(\alpha} V_{\beta)} = 0$ and hence $V$ is a
Killing vector \cite{gibbons:93}. Combining this with
\p{eqn:lie{}F} implies that $V$ is the generator of a symmetry of
the full solution $(g,{}F)$. Rewriting \p{deevee} as \be
\label{eqn:dV}
  dV = \frac{4}{\sqrt{3}} f {}F + \frac{2}{\sqrt{3}} * ({}F \wedge V)
\ee and then taking the exterior derivative gives \be
  0 = \frac{2}{\sqrt{3}} \left[ {\cal L}_V * {}F - i_V \left( d*{}F
  + \frac{2}{\sqrt{3}} {}F \wedge {}F \right) \right].
\ee The first term on the right hand side vanishes as a
consequence of equations \p{eqn:lie{}F} and the fact that $V$ is a
Killing vector. The second term vanishes\footnote{Note that this
calculation indicates the consistency between the sign of the
Chern-Simons term in the supergravity action and the sign and
factors appearing in the Killing spinor equation.} if one imposes
the equation of motion for ${}F$.

{}Finally, differentiating $X^{(i)}$ gives \be \label{eqn:dPhi}
  D_\alpha X^{(i)}_{\beta\gamma} = -\frac{1}{\sqrt{3}} \left[ 2
  {}F_{\alpha}{}^{\delta} \left( *X^{(i)} \right)_{\delta\beta\gamma} -2
  {}F_{[\beta}{}^{\delta} \left( *X^{(i)} \right)_{\gamma] \alpha
  \delta} + \eta_{\alpha [\beta} {}F^{\delta \epsilon} \left( *X^{(i)}
  \right)_{\gamma] \delta\epsilon} \right],
\ee which implies \be\label{deephi}
  d X^{(i)} = 0,
\ee and \be \label{eqn:dstarPhi}
  d * X^{(i)} = -\frac{2}{\sqrt{3}} {}F \wedge X^{(i)}.
\ee
To make further progress we will examine separately the case in which
the  Killing vector is time-like and the case in which it is null in
the two
following sections. More precisely, the case in which $f$ vanishes
everywhere will be analyzed in section \ref{sec:null}. If $f$ does not
vanish everywhere then pick a point $p$ at which $f \ne 0$. By
continuity, $f$ must be non-zero in a neighbourhood ${\cal U}$ of
$p$. The analysis of section \ref{sec:timelike} will give the general
solution in ${\cal U}$ and this can then be analytically extended to
the whole spacetime.

\sect{The timelike case}

\label{sec:timelike}

\subsection{Introduction}

In this section we shall consider the case in which $f$ is
non-zero and hence $V$ is a timelike Killing vector field.
Equation \p{eqn:XcontX} implies that the 2-forms $X^{(i)}$ are all
non-vanishing. Introduce coordinates such that $V =
\partial/\partial t$. The metric can then be written locally as
\be
\label{eqn:metric} ds^2=f^2(dt+\omega)^2-f^{-1}h_{mn}dx^m dx^n \ee
where we have assumed, essentially with no loss of generality, $f>0$
(we shall return to this point shortly).
The metric $f^{-1} h_{mn}$ is obtained by
projecting the full metric perpendicular to the orbits of $V$. The
manifold so defined will be referred to as the base space $B$ and we
will deduce that $h_{mn}$ is a hyper-K\"ahler metric.

Define
\be \label{eqn:e0def}
  e^0 = f (dt+\omega)
\ee
and if $\eta$ defines a positive orientation on $B$ then we
use $e^0\wedge \eta$ to define a positive orientation for the D=5
metric.
The two form $d\omega$ only has components tangent to the base
space and can therefore be split into self-dual and anti-self-dual
parts with respect to the metric $h_{mn}$: \be f
d\omega=G^{+}+G^{-} \ee where the factor of $f$ is included for
convenience. Equations \p{eqn:df} and \p{eqn:dV} can now be solved
for ${}F$, giving \be \label{eqn:{}Fsol}
  {}F = \sqrt{3} \left( -\frac{1}{2} f^{-2} V \wedge df + \frac{1}{6} G^+
  + \frac{1}{2} G^- \right),
\ee
which can also be written
\be
\label{eqn:{}Fsol2}
  {}F = \frac{\sqrt{3}}{2} de^0  - \frac{1}{\sqrt{3}} G^+.
\ee
Note that for all previously known solutions,
including the BMPV black hole \cite{bmpv,gauntlett:99}
(which will be briefly reviewed below),
  $d\omega$ is anti-self-dual so the second term
on the right-hand-side of  \p{eqn:{}Fsol2}
is absent. The Bianchi identity and equation of
motion for ${}F$ imply the following equations: \be
  dG^{+}=0
\ee and \be \label{eqn:gaugeeqns} \Delta
f^{-1}=\frac{4}{9}(G^{+})^2 \ee where $\Delta$ is the Laplacian in
the metric $h$, and $(G^{+})^2\equiv (1/2)(G^{+})_{mn}(G^{+}){^{mn}}$
where the indices here are raised with $h_{mn}$.

Equation \p{eqn:VdotX} implies that the $2$-forms $X^{(i)}$ can be
regarded as $2$-forms on the base space and Equation
\p{eqn:VstarX} implies that they are anti-self-dual: \be
  *_4 X^{(i)} = - X^{(i)},
\ee where $*_4$ denotes the Hodge dual associated with the metric
$h_{mn}$. Equation \p{eqn:XcontX} can be written \be
\label{eqn:quat}
  X^{(i)}{}_m{}^p X^{(j)}{}_p{}^n = - \delta^{ij} \delta_m{}^n
  + \epsilon_{ijk} X^{(k)}{}_m{}^n
\ee where indices $m,n, \ldots$ have been raised with $h^{mn}$,
the inverse of $h_{mn}$. This equation shows that the $X^{(i)}$'s
satisfy the algebra of   imaginary unit quaternions.
Furthermore, we find that   \p{eqn:dPhi} yields
\be
  \nabla_m X^{(i)}_{np} = 0,
\ee
where $\nabla$ is the Levi-Civita connection associated with
$h_{mn}$.
Combined with \p{eqn:quat} this shows that the base
space does indeed admit an integrable hyper-K\"ahler structure.

We have exhausted the content of the equations satisfied by the
bosonic quantities. We next examine the Killing spinor equation
itself and find that it imposes no further conditions. In an
orthonormal basis with time direction $e^0$ given by
\p{eqn:e0def}, equation \p{eqn:Vproj} implies \be
\label{eqn:gamma0proj}
  \gamma^0 \epsilon = \epsilon.
\ee
Using this with
\p{eqn:fivegamma} one can show that $\gamma^{ij} \epsilon$ is
anti-self-dual with respect to the base space metric and hence \be
  G^+_{ij} \gamma^{ij} \epsilon = 0.
\ee Here $i,j=1 \ldots 4$ refer to components in a basis
orthonormal with respect to $h_{mn}$ (but $\gamma^{ij}$ is still
defined in terms of the five dimensional gamma matrices).
Using these results, the $0$ component of the Killing spinor
equation then implies that $\epsilon$ is time-independent.
The spatial components are then solved if
\be
  \epsilon (t,x) = f^{1/2} \eta (x)
\ee where $\eta(x)$ is a covariantly constant spinor on the
hyper-K\"ahler base space. Now any hyper-K\"ahler space admits
covariantly constant chiral spinors satisfying $\gamma^{1234} \eta
= \eta$ if the K\"ahler-forms are anti-self dual. Noting that this
chirality condition is actually a consequence of
\p{eqn:gamma0proj} implies that \p{eqn:gamma0proj} is the only
projection imposed on the Killing spinors and we deduce that the
configurations preserve at least 1/2 of the
supersymmetry.

We have imposed the Bianchi identity for ${}F$ and the ${}F$ field
equation, so we should also check whether the Einstein equations
are satisfied. They are in fact automatically
satisfied
as one can deduce from the integrability condition for the Killing
spinor
equation presented in appendix B. {}From there we have that
\bea
E_{\mu\nu}V^\nu&=&0\nn E_{\mu\nu}E_\mu{}^\nu&=&0\qquad {\rm
no\quad sum\quad on}\quad \mu
\eea
where $E_{\mu\nu}=0$ is equivalent to the Einstein equations.
Working in the orthonormal frame, the first condition implies
$E_{0 0}=E_{i0}=0$ and the second then implies $E_{ij}=0$.

In summary, the above analysis shows that the general
supersymmetric solution in the stationary case with $f>0$ is determined
by a
hyper-K\"ahler base 4-manifold $B$ with metric $h_{mn}$ and an
orientation
chosen so that the hyper-K\"ahler two-forms are anti-self-dual,
together with a globally defined function $f$ and locally defined 1-form
connection $\omega$ on $B$. Writing $fd\omega=G^-+G^+$, we have
$dG^+=0$ and also $\Delta f^{-1}=(4/9)(G^+)^2$. The field strength
is then determined as in \p{eqn:{}Fsol}.
These solutions\footnote{Note
that given a hyper-K\"ahler metric, the equations to be solved can
be obtained from varying an action functional on $B$. To see this
we first introduce the field strength $\ff=d \omega$ and then note
that $f$ must satisfy \be d(f \ff ^+)=0, \qquad \Delta  f^{-1}
=\frac{4}{9} f^2 (\ff ^{+})^2 \ee where $\ff
^{+}=\frac{1}{2}(\ff+*_4\ff)$. These equations follow from varying
the action on $B$ \be
  S= \int _B d^4 x \sqrt{h} \left( \frac{1}{2} (\partial \sigma)^2
  -\frac{4}{9}\sigma ^{-1} (\ff ^{+})^2 \right)
\ee with respect to $\omega$ and $\sigma$, where $\sigma =f^{-1}$.}
preserve
at least 1/2 of the supersymmetry, with Killing spinors satisfying
\p{eqn:gamma0proj}.

When $f<0$ an identical analysis reveals that the most general
supersymmetric solution is simply obtained from this solution by simply
taking $t\to -t, \omega\to -\omega$
and reversing the orientation on the base manifold $B$.
In other words
\bea
ds^2&=&|f|^2(dt+\omega)^2-|f|^{-1}h_{mn}dx^m dx^n \nn
  {}F &=& -\left(\frac{\sqrt{3}}{2} de^0 -
\frac{1}{\sqrt{3}} G^-\right).
\eea
with $e^0=|f|(dt+\omega)$ and positive orientation $e^0\wedge \eta$,
where $\eta$
is a positive orientation on the base manifold,
is a supersymmetric solution with Killing spinors satisfying
$\gamma^0\epsilon=-\epsilon$
provided that $h$ is a hyper-K\"ahler metric
with self-dual hyper-K\"ahler two-forms, $|f|$ is a globally defined
function and
writing $|f| d\omega =G^++G^-$ one demands
$dG^-=0$ and also $\Delta |f|^{-1}=(4/9)(G^-)^2$.

In subsequent subsections we will construct explicit solutions
working with the case $f>0$ for definiteness.
The subsections can be essentially read independently of each other.
Subsection \ref{subsec:static} discusses static solutions;
\ref{subsec:compact} solutions
when the base space is compact; \ref{subsec:godel} the maximally
supersymmetric
G\"odel-type solution; \ref{subsec:flat} some further solutions with a
flat hyper-K\"ahler
base space including a quick review of the BMPV black hole;
\ref{subsec:eh} focuses on solutions with base space Eguchi Hanson and
Taub-NUT as
there are some similarities with the ``Resolution
Through Transgression'' papers;
\ref{subsec:gh} the general solution for the case that the base space
has a Gibbons-Hawking metric \cite{gibbons:78} and a discussion
of how the IWP solutions of $N=2$, $D=4$ supergravity can
be obtained via dimensional reduction.

\subsection{Static solutions}

\label{subsec:static}

{\bf Proposition.} The stationary Killing vector field $V$ is
hyper-surface orthogonal if, and only, if $G^-=0$.

\noindent {\it Proof.} It is easy to show that $V$ is
hyper-surface orthogonal if, and only if, $d\omega=0$. Clearly
$d\omega=0$ implies $G^-=0$. Conversely, assume $G^- = 0$. Closure
of $G^+$ then gives \be
  df \wedge d\omega = 0 \quad \Rightarrow \qquad df \wedge G^+ = 0.
\ee The dual of this gives \be
  \partial_m f G^{+mn} = 0,
\ee so that \be
  (df \wedge G^+)_{mnp} G^{+np} = 0 \qquad \Rightarrow \qquad
  \partial_m f \left( G^+ \right)^2 = 0.
\ee Hence if $G^- = 0$ then either $df=0$ or $G^+ =0$. In the
former case, the equation \p{eqn:gaugeeqns}   for $f$ implies $G^+ = 0$.
Hence $G^- = 0$ implies $G^+ = 0$ and therefore $d\omega = 0$.

It follows from this proposition that if $G^-=0$ then, at least
locally, there will be a function $\lambda(x)$ such that $\omega =
d\lambda$. A coordinate transformation $t = t' - \lambda(x)$ then
brings the metric to a manifestly static form.
The solution can then be written
\bea
  ds^2 &=& f^{2} dt^2 - f^{-1} h_{mn} dx^m dx^n, \nn
{}F&=&\frac{\sqrt{3}}{2} df\wedge dt,
\eea
where $f^{-1}$ is a harmonic function on the base space. {}For a
harmonic function with a
single pointlike source  on a flat base space this gives the
non-rotating
black hole solution.

Note that these static solutions have vanishing magnetic charge.
This does not mean that there are no supersymmetric magneto-static
solutions but simply that the static Killing vector of such a
solution cannot be written as $\bar{\epsilon} \gamma_{\alpha}
\epsilon$ with $\epsilon$ a Killing spinor. {}For example, it will
be shown in section \ref{sec:null} that the magnetic black string
solution corresponds
to the null case $f \equiv 0$.

\subsection{Solutions with compact base space}

\label{subsec:compact}

If the base space is compact then it must be K3 (with self-dual
curvature) or
$T^4$. Suppose $f$ is smooth and non-vanishing on the base space.
Integrating
equation \p{eqn:gaugeeqns} yields $G^+ = 0$ and hence $f^{-1}$ is
harmonic. However there are no non-trivial smooth harmonic
functions on a compact manifold so $f$ must be constant. By
rescaling $t$, $\omega$ and $h_{mn}$ we can set $f=1$. This leaves
\be
  d\omega = G^-,
\ee so $G^-$ is closed and anti-self-dual and therefore harmonic.
Taking the wedge product of this equation with $G^-$ and
integrating over the base space shows that if $G^-$ is non-zero
then $\omega$ cannot be globally defined.

The only anti-self-dual harmonic forms on K3 (with self-dual curvature)
or
$T^4$ are the complex structures so it follows that \be
  d\omega =  4 \godpar J,
\ee where $J$ is an anti-self-dual complex structure and $\godpar$
a constant. The field strength is
\be \label{ab1}
  {}F = 2\sqrt{3} \godpar J,
\ee
and the metric is
\be \label{ab2}
  ds^2 = (dt + \omega)^2 - h_{mn} dx^m dx^n.
\ee These solutions describe rotating closed universes containing
a constant magnetic field. The case of $T^4$ can be discussed more
explicitly. Local coordinates can be chosen such that \be \label{ab3}
  h_{mn} dx^m dx^n  = (dx^1)^2 + (dx^2)^2 + (dx^3)^2 + (dx^4)^2,
\ee and \be
  J = \left( dx^1 \wedge dx^2 - dx^3 \wedge dx^4 \right).
\ee One could then take \be\label{omegglob}
  \omega = 2 \godpar \left( x^1 dx^2 -x^2dx^1-x^3dx^4+x^4dx^3 \right).
\ee This is clearly not globally defined on $T^4$.
Rather than discuss this further we shall discuss the analogous
solution on the covering space of $T^4$, namely $\bR^4$.

\subsection{A supersymmetric analogue of the G\"odel universe}

\label{subsec:godel}

The supersymmetric G\"odel universe has metric
given by \p{ab2},\p{ab3},\p{omegglob} and gauge field given by \p{ab1}.
For simplicity we let $\godpar=1/4$.
On the base space $\bR^4$, write $x^1 = r_1\cos \phi_1$,   $x^2 =
r_1\sin \phi_1$, $x^3 = r_2\cos \phi_2$, $x^4
= r_2\sin\phi_2$ and using $\omega$ as in \p{omegglob}
the solution can be written\footnote{ This solution has previously
appeared in \cite{tseytlin:96} as a footnote.}
\bea\label{godexplicit}
  ds^2 &=& \left( dt +\frac{1}{2}(r_1^2 d\phi_1 - r_2^2 d\phi_2)
  \right)^2 - \left(dr_1^2 + r_1^2 d\phi_1^2 + dr_2^2 + r_2^2 d\phi_2^2
  \right)\nn
{}F&=&\frac{\sqrt{3}}{2}(r_1dr_1\wedge d\phi_1
-r_2dr_2\wedge d\phi_2).
\eea
The metric has coordinate singularities at
$r_1 = 0$ and $r_2 = 0$ but these can be removed by going back to
Cartesian coordinates. Note that $\partial/\partial \phi_i$
is timelike for $r_i >2$ so this solution has closed timelike
curves in those regions. Note that the signature remains Lorentzian.
This is
very similar to what happens in the four dimensional G\"odel universe
\cite{godel}. There are further similarities between our solution and
G\"odel's. The G\"odel solution is defined on a manifold of topology
$\bR^4$.
Ours has topology $R^5$. The matter content of G\"odel's solution
consists
of pressureless dust balanced by a negative cosmological constant.
Calculating the energy-momentum tensor  for the $F_{\mu\nu}$ of our
solution
one finds that it (and all the solutions of the previous section) has
vanishing pressure and constant energy density proportional to
$\godpar^2$, i.e., the electromagnetic field has the same
energy-momentum  as
  pressureless dust. In addition, just like the G\"odel universe, this
solution is homogeneous;
it was shown in \cite{Alonso-Alberca:2002wr} that the near-horizon
geometry of the
BMPV black hole is also homogeneous.
It is straightforward to verify that the following
vectors in cartesian co-ordinates are Killing
vectors:
\bea
&&V=\frac{\partial}{\partial t},\nn
B_1=\frac{\partial}{\partial x^1}
-\frac{x^2}{2}\frac{\partial}{\partial t},&&
\qquad
B_2=\frac{\partial}{\partial x^2}
+\frac{x^1}{2}\frac{\partial}{\partial t},\nn
B_3=\frac{\partial}{\partial x^3}
+\frac{x^4}{2}\frac{\partial}{\partial t},&&
\qquad
B_4=\frac{\partial}{\partial x^4}
-\frac{x^3}{2}\frac{\partial}{\partial t},\nn
R_1=x^1\frac{\partial}{\partial x^2}-x^2\frac{\partial}{\partial
x^1},&&\qquad
R_2=x^3\frac{\partial}{\partial x^4}-x^4\frac{\partial}{\partial x^3}
\eea
which act transitively. The last two Killing-vectors generate a
$U(1)\times U(1)$ group of rotations in $\bR^4$. In fact, as we will
see later (see equation \p{godsolution})
this is actually enlarged to an a $SU(2)\times U(1)$
group of rotations, giving a $9$ parameter family of isometries, the
same number as for $AdS_2 \times S^3$ or $AdS_3 \times S^2$.

Surprisingly, this solution is maximally supersymmetric preserving
all 8 supersymmetries of the theory. Explicitly, using the obvious
frame $(e^0,e^i)=(dt+\omega,dx^i)$, the Killing spinors are given by
\be
\epsilon=\theta^+ + (1+J_{ij}x^i\gamma^j)\theta^-
\ee
where $\theta^\pm$ are constant spinors satisfying
$\gamma^0\theta^\pm =\pm\theta^\pm$.

We can determine the symmetry superalgebra using the
method of \cite{Gauntlett:1998kc} (see also \cite{Townsend:1998ci,
Figueroa-O'Farrill:1999va,Figueroa-O'Farrill:2001nz}).
This is a Lie-superalgebra whose even subspace ${\cal B}$
is spanned by the above Killing-vectors and the
odd subspace ${\cal F}$ by the Killing spinors.
The bilinear map ${\cal B}\times {\cal B}\to {\cal B}$ is the
Lie-bracket of the above vector fields. The map
${\cal B}\times {\cal F}\to {\cal F}$ is obtained from the
Lie-derivative of the Killing spinors with respect to the Killing
vectors
and the map ${\cal F}\times {\cal F}\to {\cal B}$ is deduced
from the Killing-vectors obtained by squaring the Killing spinors.
It is straightforward to write all of these maps out explicitly, but we
shall
just record the last map. We first note that for any two Killing
spinors $\epsilon$, $\epsilon'$ we can construct a Killing vector via
$K^\mu=\bar\epsilon\gamma^\mu \epsilon'$.
If we let $\epsilon=\theta^+ + (1+J_{ij}x^i\gamma^j)\theta^-$ and
$\epsilon'=\rho^+ + (1+J_{ij}x^i\gamma^j)\rho^-$ we find
\bea
\bar\epsilon\gamma^\alpha \epsilon' E_\alpha
=\overline{\theta^+}\rho^+ V
+\overline{\theta^-}\rho^- (-V-2R_1+2R_2)
+(\overline{\theta^+}\gamma^i\rho^-
+\overline{\theta^-}\gamma^i\rho^+) B_i
\eea
where $E_\alpha$ are the vector fields dual to the frame introduced
above.

\subsection{Further solutions with flat base-space}

\label{subsec:flat}

Let us now present some additional new solutions with base
space $\bR^4$ that admit at least an $SU(2)$ sub-group of isometries
of the $SO(4)$ rotation group of $\bR^4$.
It will be useful to work with $SU(2)$
Euler-angles, which we introduce via
\bea
x^1+ix^2&=&r\cos\frac{\theta}{2}e^{i(\frac{\psi+\phi}{2})}\nn
x^3+ix^4&=&r\sin\frac{\theta}{2}e^{i(\frac{\psi-\phi}{2})}
\eea
with $0 \leq \theta < \pi$, $0 \leq \phi < 2 \pi$ and $0 \leq \psi
<4 \pi$. We also work with left-invariant one-forms $\sigma_R^i$
satisfying $d \sigma_R^i = {1/2} \epsilon^{ijk} \sigma_R^j
\wedge \sigma_R^k$ and right-invariant one-forms $\sigma_L^i$
satisfying $d \sigma_L^i = -{1/2} \epsilon^{ijk} \sigma_L^j
\wedge \sigma_L^k$. The subscripts refer to the fact that, for
example, the left invariant one-forms $\sigma^i_R$ are dual to right vector
fields that generate right actions. Explicit expressions can be
found in appendix A. A positive
orientation is fixed by $dx^1\wedge dx^2\wedge dx^3\wedge dx^4=
dr\wedge(\frac{r}{2})\sigma^1_R\wedge(\frac{r}{2})\sigma^2_R\wedge(\frac
{r}{2})
\sigma^3_R=
dr\wedge(\frac{r}{2})\sigma^1_L\wedge(\frac{r}{2})\sigma^2_L\wedge(\frac
{r}{2})
\sigma^3_L$. The flat metric on $\bR^4$ is given by
\be
ds^2=dr^2+\frac{r^2}{4}[(\sigma^1)^2+(\sigma^2)^2+(\sigma^3)^2]
\ee
for either left or right invariant one-forms.

Let us begin by recording the rotating BMPV solution
\cite{bmpv,gauntlett:99}.
We write the flat base using left-invariant one-forms $\sigma^i_R$.
As previously noted it has $G^+=0$ and is given by:
\bea
\label{eqn:bmpvsol}
f^{-1}&=&1+\frac{\mu}{r^2}\nn
\omega&=&\frac{j}{2r^2}\sigma^3_R
\eea
The ADM mass and angular momentum are given by
\bea
M&=&\frac{3\pi\mu}{4 G}\nn J&=&-\frac{j\pi}{2G}
\eea
This angular momentum corresponds to equal rotation in the 1-2 and 3-4
planes
with opposite sign. Recall that all closed timelike curves are hidden
behind the
horizon at $r=0$ providing that $|j|\le \mu^{3/2}$.

A simple generalization of the BMPV black hole solution
with $G^+\ne 0$ can be obtained by considering the general ansatz
\be\label{gummy}
  \omega  = \Psi (r) \sigma_R^3 \ee
and assuming $f=f(r)$.
Demanding that $G^+$ is closed implies that $f(r\Psi'+2\Psi)=\chi r^2$
for constant $\chi$. Solving \p{eqn:gaugeeqns} we then find
\bea
f^{-1}&=&\lambda+\frac{\mu}{ r^2}+\frac{\chi^2}{9}r^2\nn
\Psi &=& \frac{j}{2r^2}
+\frac{\chi\mu}{2}+\frac{\chi\lambda}{4}r^2+\frac{\chi^3}{54}r^4
\eea
for constant $\lambda,\mu, j$.
This solution has
\bea
G^+=\frac{\chi}{4}d(r^2\sigma_R^3)
=\chi(dx^1\wedge dx^2+dx^3\wedge dx^4)
\eea
Supposing $\lambda\ne0$ we can always choose
$\lambda=1$ by rescaling the radial and time co-ordinates. We note
that the solution will have closed time-like curves when
$\Psi^2f^2-f^{-1}r^2/4$ is positive.
When $\chi=0$ we return to the rotating black hole solution.
When $\chi\ne 0$, the solution is no longer asymptotically flat and
is a rotating   universe with both electric and magnetic fields and with
closed time-like curves.

{}Further new solutions with flat base space can be found by working
with right
invariant 1-forms on $SU(2)$, $\sigma^i_L$. We now look for solutions
with
\be\label{gum}
\omega=\Psi(r) \sigma^3_L
\ee
and $f=f(r)$. {}Following the same steps as above we find the solution
\bea
f^{-1} &=& \lambda+ \frac{\mu}{r^{2}}+ {\chi^2
\over 27 r^6} \nn
\Psi (r) &=& \godpar r^2 - \chi
(\frac{\lambda}{4 r^{2}} +{\mu \over 6r^4} +{\chi^2 \over 270 r^8})
\eea
where $\lambda,\mu,\chi$ are again constant. {}For this solution $G^+$
is given by
\be G^+
=-\frac{\chi}{4}d\left[ \frac{1}{r^{2}}\sigma^3_L\right]
\ee
Let us now discuss this solution in more detail.
We first restrict to
$\lambda=1$ which can be achieved by rescaling the radial and time
co-ordinates when $\lambda\ne0$.
It appears that these solutions are non-singular at $r=0$. {}For
example, if one calculates ${}F^2$, which from  Einstein's
equations is proportional to the Ricci scalar, one finds for
$\chi\ne 0$ it goes to zero like $r^4$, while for $\chi=0$ it goes
to a constant. Other curvature invariants also appear to be regular at
$r=0$.

Note also that if we set $\mu=\chi=0$ and $\gamma \neq 0$, then the
solution
is given by
\bea\label{godsolution}
ds^2&=&(dt+\gamma r^2 \sigma^3_L)^2-(dr^2+\frac{r^2}{4}
[(\sigma^1_L)^2+(\sigma^2_L)^2+(\sigma^3_L)^2])\nn
F&=&\frac{{\sqrt 3}\gamma}{2}d(r^2\sigma^3_L)
\eea
which is just the generalised G\"odel solution introduced above. We
already have shown
that the solution has a 7 parameter family of isometries including
$U(1)\times U(1)$ rotations in $\bR^4$. In the co-ordinates of this
section, it is clear that the $U(1)\times U(1)$ symmetry is
enlarged to $SU(2)\times U(1)$ corresponding to right $SU(2)$
actions and a left $U(1)$ action (in the 3 direction).

Alternatively, we may set $\gamma=0$, $\mu \neq 0$, $\chi \neq 0$.
The metric is then asymptotically flat. The ADM mass and
angular momentum are given by
\bea
M&=& {3 \pi \mu \over 4 G}\nn J&=&\frac{\chi\pi}{4G}
\eea
Here the angular momentum corresponds to equal rotation in the 1-2 and
3-4
plane since $d\omega$ is self-dual\footnote{One can get a solution with
the same quantum numbers as the black hole solution \p{eqn:bmpvsol}
by taking $t\to-t, \phi\leftrightarrow \psi$. This corresponds to
switching to a solution
with $f<0$, as described at the end of section 3.1.}.
{}Furthermore, it is clear from examining the sign of $ f^2
\Psi^2-{1 \over 4} r^2 f^{-1}$ that there are closed timelike
curves for all values of $\mu>0$ within $0<r<r_{\rm crit}$
provided that $\chi \neq 0$. By analysing the behaviour of massive test
particles it appears that this geometry is geodesically complete
and that $r=0$ exhibits a repulson behaviour. Moreover by tuning
the angular momentum of the test particle it can
approach $r=0$ arbitrarily closely and thus enter the time-machine.
The global causal structure of this spacetime is similar to that of
the over-rotating BMPV solution presented in \cite{Gibbons:1999uv}.
However, unlike the general family of rotating BMPV solutions,
tuning $M$ and $J$ in the region $M>0$, $J>0$ does not alter the
causal structure of the spacetime.

There are a number of straightforward generalisations of the
new solutions we have presented here. For example, we could replace
\p{gummy},\p{gum} by $\Psi_i(r)\sigma^i$.

\subsection{Rotating Eguchi-Hanson and Taub-NUT}

\label{subsec:eh}

When $G^+=0$ the function $f^{-1}$ is harmonic on the hyper-K\"ahler
base. When $G^+\ne 0$ it is modified by the square of a closed,
self-dual and hence harmonic form via equation \p{eqn:gaugeeqns}.
This is reminiscent of the equations arising in the ``Resolution
Through Transgression'' papers (for a review see
\cite{cveticetal}). In particular an interesting regular
generalisation of both the Eguchi Hanson space and Taub-NUT space
with flux was constructed in section 3 of \cite{Cvetic:2000mh}.
We therefore examine the Eguchi-Hanson and the Taub-NUT cases in
more detail, finding regular supersymmetric solutions, albeit with
closed time-like curves.

Let us first consider the Eguchi-Hanson case. The metric is given
by
\bea\label{eguchih}
ds^2=W^{-1}dr^2+{r^2\over 4}((\sigma^1_L)^2+(\sigma^2_L)^2)
+W{r^2\over 4}(\sigma^3_L)^2
\eea
where \be W=1-{a^4\over r^4} \ee
This is a regular space provided that the range of
$\psi$ is $0\le\psi\le 2\pi$ and of
$r$ is $a\le r\le \infty$ and $r=a$ is the $S^2$ bolt.
We choose positive orientation to be
given by $dr\wedge\sigma^1_L \wedge\sigma^2_L\wedge\sigma^3_L$ so
that the three K\"ahler forms are anti-self-dual. If
$\omega=0$ then the harmonic function $f^{-1}$ is constant or singular.
The singular solutions can be resolved in the following sense.
We choose
\be\label{pos}
G^+=-{\chi\over 4}d(r^{-2}\sigma^3_L) \ee
and hence need to solve
\be \Delta
f^{-1}={8\chi^2\over 9r^8} \ee where $\Delta$ is the Laplacian
with respect to the Eguchi-Hanson metric, and one should note
a key sign difference with the similar equation in \cite{Cvetic:2000mh}.
The general solution is given by
\be
f^{-1}=\lambda-{\chi^2\over 9a^4 r^2} + \delta \log
\frac{(r^2-a^2)}{(r^2+a^2)}
\ee
where $\lambda,\delta$ are arbitrary integration constants.
If we seek solutions of the form
\be\label{anty}
\omega=\Psi(r)\sigma^3_L \ee
we find
\be
\Psi=-{\chi\lambda\over 4r^2}+{\chi^3\over 54 r^4 a^4}+
\frac{\delta \chi}{4r^2a^4}[(r^4-a^4)\log
\frac{(r^2-a^2)}{(r^2+a^2)}+2a^2r^2]
+\godpar r^2
\ee
A regular solution can be obtained by first setting setting $\delta=0$
to get
\bea\label{bing}
f^{-1}&=&\lambda-{\chi^2\over 9a^4 r^2} \nn
\Psi&=&-{\chi\lambda\over 4r^2}+{\chi^3\over 54 r^4 a^4}+\godpar r^2
\eea
Restricting to $\chi^2\le
\lambda 9a^6$, $f^{-1}$ will be non-zero for $a\le r\le \infty$.
By calculating $F^2$ it appears that this five-dimensional
solution is regular, provided that we restrict $\chi$ as
stated. To eliminate closed time-like curves as $r\to \infty$
we set $\godpar=0$. However, it seems that there are always closed
time-like curves near $r=a$.

It is interesting to note that we can find analogous solutions if we
start with a singular hyper-K\"ahler space given by \p{eguchih} with
$W=1+b^4/r^4$. In this case there is no reason to take $\phi$ or
$\psi$ to have particular ranges. If we again choose
\p{pos},\p{anty} the solution is given by
\bea\label{lasty}
f^{-1}&=&\lambda+{\chi^2\over 9b^4 r^2} + \delta
\arctan\frac{r^2}{b^2}\nn
\Psi&=&-{\chi\lambda\over 4r^2}-{\chi^3\over 54 r^4 b^4}-
\frac{\delta \chi}{4r^2b^4}[(r^4+b^4) \arctan\frac{r^2}{b^2} + b^2 r^2]
+\godpar r^2
\eea
These solutions appear to be singular in general. Note that if we take
$0 \le \phi \le 2\pi$ and $0 \le \psi \le 4\pi$ then the Eguchi-Hanson
space is asymptotically Euclidean and the five dimensional solution is
asymptotically flat if the constants of integration are chosen
appropriately.

Surprisingly, if we set $\delta=\lambda=\gamma=0$ we obtain
the maximally supersymmetric $AdS_2\times S^3$ solution as we will
show in section \ref{sec:maximal}.

Let us now consider the base space to be Taub-NUT space. The
metric is given by \be
ds^2=\frac{(r+a)}{(r-a)}dr^2+(r^2-a^2)(
(\sigma^1_R)^2+(\sigma^2_R)^2)+ 4a^2\frac{(r-a)}{(r+a)}(\sigma^3_R)^2
\ee
When $a$ is positive,
the range of $\psi$ is $0\le\psi\le 4\pi$, and that of $r$ is
$a\le r\le\infty$, so that the topology of the space is $\bR^4$. If
we take positive orientation to be given by
$adr\wedge\sigma^1_R\wedge\sigma^2_R\wedge\sigma^3_R$
then the hyper-K\"ahler forms are
anti-self-dual. We now choose \be G^+=\chi
d\left[ \frac{(r-a)}{(r+a)}\sigma^3_R\right] \ee and hence \be
\Delta f^{-1}={8\chi^2\over 9(r+a)^4} \ee where $\Delta$ is the
Laplacian with respect to the Taub-NUT metric. The solution to this is
given by
\be f^{-1}=\lambda-{2\chi^2\over
9a(r+a)} +\frac{\delta}{r-a}
\ee
As above we
let
\be
\omega=\Psi(r)\sigma^3_R\ee
and find
\be
\Psi(r)=-\frac{16\chi^3a}{27(r+a)^2(r-a)}+\frac{2\chi(2\chi^2-9 \delta
a +
18\lambda a^2)}{9(r+a)(r-a)} -\frac{4\chi\lambda a}{(r-a)}+\gamma
\frac{(r+a)}{(r-a)}
\ee
In order to
construct a regular solution, we set $\delta=0$ and
choose $\gamma=\chi \lambda-\chi^3/27a^2$ to get
\bea\label{bad}
  f^{-1}&=&\lambda-{2\chi^2\over
9a(r+a)}\nn
\Psi&=& {\chi (r-a)( 27 \lambda a^2 (r+a) - \chi^2 (r+5 a))
\over 27 a^2 (r+a)^2}
\eea
Then $f^{-1}$ will be non-zero for $a\le r\le \infty$ and note
that $\Psi(a)=0$, indicating the
regularity of the solution. It is straightforward to check that
$F^2$ is also regular everywhere, providing that we restrict
$\chi$ as mentioned. Again this solution has closed time-like
curves.

Once again we can build solutions from the singular
negative mass Taub-NUT space by letting $a$ be negative.
For example, if we take \p{bad} with
$\lambda=0$ we will show in section \ref{sec:maximal} that we get
a maximally supersymmetric solution that is in fact the G\"odel
solution.

\subsection{Solutions with Gibbons-Hawking base space}

\label{subsec:gh}

In this subsection, the equations for $f$ and $\omega$ will be
examined in more detail for the case of a Gibbons-Hawking base space.
The solutions of the last three subsections will comprise special cases.
It has been shown \cite{gibbons:88} that if a four dimensional
hyper-K\"ahler manifold admits a triholomorphic Killing vector
field, that is, a Killing vector field $L$ that preserves the
complex structures (${\cal L}_L X^{(i)} = 0$), then it must be a
Gibbons-Hawking \cite{gibbons:78} metric: \ba
  ds^2 &=& H^{-1} \left( dx^5 + \chi_i dx^i \right)^2 + H dx^i dx^i, \\
  \nabla \cross \bchi &=& \nabla H. \nonumber
\ea The Killing vector field is $\partial/\partial x^5$. $\nabla$
is the flat connection on the Euclidean 3-space with coordinates
$x^i$ and $H$ is harmonic on this space. The complex structures
are given by \cite{gibbons:88} \be
  X^{(i)} = \left( dx^5 + \chi_j dx^j \right) \wedge dx^i - \frac{1}{2}
  H \epsilon_{ijk} dx^j \wedge dx^k.
\ee Anti-self-duality of these forms fixes the orientation of the
base space so that the volume form is \be
  H dx^5 \wedge dx^1 \wedge dx^2 \wedge dx^3.
\ee
Examples of Gibbons-Hawking metrics are: flat space ($H=1$ or
$H=1/|{\bf x}|$), Taub-NUT space ($H=1+2M/|{\bf x}|$) and the
Eguchi-Hanson space ($H=2M/|{\bf x}| + 2M/|{\bf x-x_0}|$)
\cite{gibbons:88}.

If the Killing vector $\partial/\partial x^5$ is a Killing
vector of the full five dimensional spacetime (i.e. if $f$ and
$\omega$ are independent of $x^5$) then the equations for $f$ and
$\omega$ can be solved explicitly. Write
\be
  \omega = \omega_5 \left(dx^5 + \chi_i dx^i \right) + \omega_i dx^i,
\ee
and introduce an orthonormal basis
\be
  e^5 = H^{-1/2} \left(dx^5 + \chi_i dx^i \right), \qquad e^i = H^{1/2}
  dx^i.
\ee
Then
\be
  G^{\pm} = -\frac{1}{2} A^{\pm}_i e^5 \wedge e^i \mp \frac{1}{4}
  \epsilon_{ijk} A^{\pm}_k e^i \wedge e^j,
\ee
where
\be
\label{eqn:Adef}
  {\bf A^{\pm}} = H^{-1} f \left[ H \nabla \omega_5 \mp \omega_5 \nabla
H \mp
\nabla \cross \bomega \right].
\ee
In equations such as this written in three-dimensional vector
notation, $\bomega$ refers to the three-vector with components
$\omega_i$. Closure of $G^+$ reduces to
\be
  \nabla \cross {\bf A^+} = 0,
\ee and \be
  \nabla \cdot (H {\bf A^+} + {\bchi} \cross {\bf A^+} ) = 0.
\ee The first of these yields \be
  {\bf A^+} = \nabla \rho,
\ee for some locally defined function $\rho$. Substituting into
the second gives
\be
  \nabla^2 (H\rho) = 0,
\ee
and hence
\be
  \rho = 3KH^{-1}
\ee
for some harmonic function $K$. The equation for $f$ reduces
to
\be
  \nabla^2 f^{-1} = \frac{2}{9} H \left( \nabla \rho \right)^2 =
  \nabla^2 \left(K^2 H^{-1} \right),
\ee
and hence
\be
  f^{-1} = K^2 H^{-1} + L,
\ee
where $L$ is another harmonic function. It remains to solve
for $\omega_5$ and $\omega_i$. Substituting the above results into
\p{eqn:Adef} gives
\be
\label{eqn:omegaeq}
  H \nabla \omega_5 - \omega_5 \nabla H - \nabla \cross \bomega
  = 3 \left(K^2 + LH \right)
\nabla \left( KH^{-1} \right).
\ee
Taking the divergence of this
gives the integrability condition
\be \nabla^2 \omega_5 = 3H^{-1}
\nabla \cdot \left[ \left(K^2 + LH \right) \nabla \left( KH^{-1}
\right) \right] = \nabla^2 \left( H^{-2} K^3 + \frac{3}{2} H^{-1}
KL \right),
\ee
with solution
\be
  \omega_5 = H^{-2} K^3 + \frac{3}{2} H^{-1} KL + M,
\ee
where $M$ is an arbitrary harmonic function. Substituting the
solution back into \p{eqn:omegaeq} then gives an equation that
determines  $\bomega$ up to a gradient (which can be absorbed
into $t$). The above analysis yields the general solution for
which the base space admits a tri-holomorphic Killing vector field
that extends to a Killing vector field of the five dimensional
spacetime. It is specified by four arbitrary harmonic functions
$H$, $K$, $L$, and $M$. Solutions with $G^+ \ne 0$ are much more
complicated than those with $G^+ = 0$ (i.e. $K \propto H$), which are
specified by three harmonic functions $H$, $f^{-1}$ and $\omega_5$.

It is worth remarking that the same solution can be derived
under rather weaker assumptions, namely that there exists a spacelike
Killing
vector field of the five dimensional spacetime that commutes with $V$
and leaves the three complex structures invariant.

The solutions with a flat base space that were discussed above can be
easily recovered in this framework. Introduce spherical polar
coordinates $(R,\theta,\phi)$ on the three dimensional flat part of
the metric. Choosing
\be H = \frac{1}{R}, \qquad {\bf \chi} =  \cos \theta d\phi,
\ee
gives a flat base space. Let $R=r^2/4$ and $x^5 = \psi$
and the metric takes the form
\be
  ds^2 = dr^2 + \frac{r^2}{4} \left( d\psi^2 + d\phi^2  + 2 \cos \theta
  d\psi d\phi + d\theta^2 \right).
\ee
The coordinates $(\theta,\phi,\psi)$ are Euler angles on
$S^3$. The solutions constructed using the left-invariant forms on
$SU(2)$ have $\omega = \Psi(r) \left(d\psi + cos \theta d\phi \right)$
and hence $\omega_5 =
\Psi$, $\bomega=0$. The full solution is obtained by taking all of the
harmonic functions to be spherically symmetric. The solutions
constructed using the right-invariant forms have $\omega = \Psi(r)
\left(d\phi + \cos \theta d\psi \right)$ and hence $\omega_5 =
\Psi(r) \cos \theta$ and $\omega_i dx^i = \Psi(r) \sin^2 \theta
d\phi$. An example is the Go\"del solution, which has $G^+=0$ and is
therefore specified by three harmonic functions: $f^{-1}=1$, $H=1/R$,
$\omega_5 \propto R \cos \theta$.

The form of the Gibbons-Hawking metric lends itself naturally to
dimensional reduction, so the above solution
yields a large class of solutions of the theory obtained by KK
reduction of the minimal five dimensional supergravity theory,
namely $N=2$, $D=4$ supergravity coupled to a vector
multiplet. It is interesting to see how the solutions of pure
$N=2$, $D=4$ can be embedded in the five dimensional theory
(for {\it maximally} supersymmetric solutions, this was done in
\cite{meessen2}). To this
end, consider a general five dimensional metric admitting a spacelike
Killing vector field $\partial/\partial x^5$ and write the metric as
\be
\label{eqn:KKans}
  ds^2 = e^{\alpha \phi} ds^2_R
  - e^{\beta \phi} \left( dx^5 + {\cal A}\right)^2,
\ee
where $ds^2_R$ is the line element of the four dimensional Lorentzian
metric
and ${\cal A}$ is a one-form potential on the reduced space.
The constants $\alpha$ and $\beta$ are chosen such that the reduced
metric is in the
Einstein frame, with a canonically normalized scalar $\phi$. Write the
five dimensional vector potential as
\be
  A = A' + \theta dx^5.
\ee
where $A'$ is another one-form potential on the reduced space.
The reduced theory has two scalars $\phi$ and $\theta$ and two one-forms
$A'$ and ${\cal A}$. We want to truncate to get the pure $N=2$,
$D=4$ theory, so we have to set these scalars to zero. Consistency
requires that their equations of motion are satisfied, which gives
\be
  \frac{3}{4} *_4 G \wedge G + *_4 F' \wedge F' = 0,
\ee
and
\be
  \frac{\sqrt{3}}{2} *_4 F' \wedge G - F' \wedge F' = 0,
\ee
where $G = d{\cal A}$ and $F'=dA'$. The orientation $\eta$ of the
reduced
spacetime has been chosen such that $dx^5 \wedge \eta$ is {\it
negatively}
oriented in five dimensions because this is what happens for our
Gibbons-Hawking solutions. These equations are both satisfied by
choosing
\be
\label{eqn:GF}
  G = -\frac{2}{\sqrt{3}} *_4 F',
\ee
which also eliminates the vector ${\cal A}$ as an independent field
(its equation of motion is satisfied using the Bianchi identity for
$F'$). Finally we are left with a theory whose equations of motion can
be derived from the action
\be
\label{eqn:4daction}
  S_4 = \frac{1}{4 \pi G_4} \int \left( -\frac{1}{4} R_4 *_4 1 -
  \frac{2}{3} *_4 F' \wedge F' \right),
\ee
which is indeed the action for the bosonic sector of pure $N=2$, $D=4$
supergravity.
To summarize, a five dimensional solution with metric of the form
\p{eqn:KKans} can be reduced to give a solution of $N=2$,
$D=4$ supergravity provided $\phi=0$, $\theta=0$ (equivalently,
$F_{\mu 5}=0$) and equation \p{eqn:GF} is satisfied.

Let's apply this to our solutions with Gibbons-Hawking base
space. The metric can be written
\be
  ds^2 = \Lambda^{-1} f H^{-1} \left( dt + \omega_i dx^i \right)^2 -
f^{-1} H dx^i dx^i - \Lambda \left(dx^5 + \chi_i dx^i - \frac{f^2
\omega_5}{\Lambda} \left(dt + \omega_i dx^i \right) \right)^2,
\ee
where
\be
  \Lambda = f^{-1} H^{-1} - f^2 \omega_5^2.
\ee
It can be verified that all of the consistency conditions for the
dimensional reduction are satisfied if we choose the harmonic
functions $L$ and $M$ such that
\be
  f^{-1} = \frac{K^2}{H}+H, \qquad \omega_5 = \frac{K}{H}f^{-1},
\ee
and
\be
  \nabla \cross \bomega = 2K \nabla H - 2H \nabla K.
\ee
The reduced metric can be written
\be
  ds_4^2 = |V|^2 \left(dt + \omega_i dx^i \right) - |V|^{-2} dx^i dx^i,
\ee
where
\be
  V^{-1} = H + iK,
\ee
$\bomega$ is given by
\be
  \nabla \cross \bomega = i \left( \bar{V}^{-1} \nabla V^{-1} - V^{-1}
\nabla \bar{V}^{-1} \right),
\ee
and the field strength is
\be
  F' = F = -\frac{\sqrt{3}}{4} \nabla_i \left( V + \bar{V} \right)
\tilde{e}^0 \wedge \tilde{e}^i - \frac{\sqrt{3}}{8} i \epsilon_{ijk}
\nabla_k \left(V - \bar{V} \right) \tilde{e}^i \wedge \tilde{e}^j,
\ee
where $\tilde{e}^0 = |V| \left(dt+\omega_i dx^i\right)$ and
$\tilde{e}^i = |V|^{-1} dx^i$ is an orthonormal basis for the four
dimensional metric. This is precisely the form of the IWP metric given
in
\cite{Tod}. So the entire timelike class of supersymmetric solutions
of the $N=2$, $D=4$ theory can be obtained by reduction of a
subset of our
solutions with Gibbons-Hawking base space. Note that the four
dimensional metric is static if ${\bomega} = 0$, which requires $K
\propto H$, which is true if, and only if, $G^+ = 0$. For example,
setting $K=0$ gives ${\bomega} = \omega_5 = 0$, $f^{-1} = H$ and the
five
dimensional metric is
\be
  ds^2 = H^{-2} dt^2 - H^2 dx^i dx^i - \left(dx^5 + \chi_i dx^i
  \right)^2,
\ee
which give the electrostatic Majumdar-Papapetrou solutions in four
dimensions.
Taking $H=1/|{\bf x}|$ gives a flat base space and the four dimensional
metric is $AdS_2 \times S^2$ (the five dimensional metric is $AdS_2
\times S^3$). $H=1+1/|{\bf x}|$ gives a Taub-NUT base
space and the four dimensional metric is extremal
Reissner-Nordstrom. Multi-centre Taub-NUT gives multi-centre
Reissner-Nordstrom. Eguchi-Hanson gives a two centre $AdS_2 \times
S^2$ solution, and multi-centre Eguchi-Hanson gives multi-centre
$AdS_2 \times S^2$. Taking $K$ to be a non-vanishing multiple of $H$
just corresponds to a duality rotation of the four dimensional gauge
field.

\sect{The null case}

\label{sec:null}

\subsection{The general solution}

In this section we shall find all solutions of minimal $N=1$, $D=5$
supergravity for which the function $f$ introduced in section 2
vanishes everywhere.

{}First introduce coordinates as follows. {}From \p{eqn:dV} it can
be seen that $V$ satisfies $V\wedge dV=0$ and is therefore
hypersurface-orthogonal. Hence there exist functions $u$ and $H$
such that \be \label{eqn:udef}
  V = H^{-1} du.
\ee A second consequence of \p{deevee} is \be
  V \cdot D V = 0,
\ee so $V$ is tangent to affinely parametrized geodesics in the
surfaces of constant $u$. One can choose coordinates $(u,v,y^m)$,
$m=1,2,3$, such that $v$ is the affine parameter along these
geodesics, and hence \be V = \frac{\partial}{\partial v}. \ee The
metric must take the form: \be
  ds^2= H^{-1} \left( {\cal {}F} du^2 + 2 du dv \right)-H^2
  \gamma_{mn} \left(dy^m + a^m du \right) \left(dy^n + a^n du \right),
\ee where the quantities $H$, ${\cal {}F}$, $\gamma_{mn}$ and
$a^m$ depend on $u$ and $y^m$ only (because $V$ is Killing). Note
that there is a lot of gauge freedom remaining. {}For example, a
coordinate transformation of the form $y \rightarrow y'(u,y)$
could be used to eliminate $a^m$. However, this freedom will be
more useful shortly.

Equations \p{eqn:VdotX} and \p{eqn:VstarX} imply that $X^{(i)}$
can be written \be
  X^{(i)} = X^{(i)}_m du \wedge dy^m.
\ee Closure of $X^{(i)}$ then gives \be
  \partial_{[m} X^{(i)}_{n]} = 0,
\ee and hence locally there exist functions $x^i(u,y)$ such that
\be
  X^{(i)}_m = \partial_m x^i.
\ee Now we can exploit the freedom to do a coordinate
transformation $y \rightarrow y'(u,y)$ by choosing $x^i(u,y)$ as
our new coordinates. The metric takes the same form as above but
with $x^i$ replacing $y^m$. In these coordinates, \be
\label{eqn:xidef}
  X^{(i)} = du \wedge dx^i.
\ee Equation \p{eqn:XcontX} now gives \be
  \gamma_{ij} = \delta_{ij},
\ee so in these coordinates, the surfaces of constant $u$ and $v$
are flat. The full metric can be written \be
  ds^2= H^{-1} \left( {\cal {}F} du^2 + 2 du dv \right)-H^2 (d{\bf x} +
  {\bf a} du)^2,
\ee where bold letters denote three dimensional quantities, e.g.,
$({\bf a})_i \equiv a^i$, with no distinction between ``up'' and
``down'' indices for such objects.

It is convenient to introduce a null basis of $1$-forms
as follows \be\label{milo}
  e^+ = V = H^{-1} du, \qquad e^- = dv + \frac{1}{2} {\cal {}F} du,
\qquad e^i = H \left(dx^i + a^i du \right). \ee These obey the
orthogonality relations \be \label{eqn:innprod}
  e^\alpha \cdot e^{\beta} = \eta^{\alpha\beta},
\ee where $\eta^{+-} = \eta^{-+} = 1$, $\eta^{ij} = - \delta_{ij}$
and other components vanish. We also choose $\epsilon_{+-123}=1$.

Equation \p{eqn:df} implies \be
  {}F = {}F_{+i} e^+ \wedge e^i + \frac{1}{2} {}F_{ij} e^i \wedge e^j.
\ee Substituting into equation \p{eqn:dV} gives \be
  {}F_{ij} = -\frac{\sqrt{3}}{2} H^{-2} \epsilon_{ijk} \nabla_k H,
\ee where $\nabla_k \equiv \partial/\partial x^k$. Equation
\p{eqn:dPhi} gives \be
  {}F_{+i} = - \frac{1}{2\sqrt{3}} H \epsilon_{ijk} \nabla_j a_k.
\ee Hence the field strength is
\be
\label{eqn:nullFsol}
  {}F = - \frac{H^{-2}}{2 \sqrt{3}} \epsilon_{ijk} \nabla_j \left( H^3
a_k
  \right) du \wedge dx^i - \frac{\sqrt{3}}{4} \epsilon_{ijk} \nabla_k
  H dx^i \wedge dx^j.
\ee The Bianchi identity reduces to \be
  \nabla^2 H = 0,
\ee \be \label{eqn:aeq}
  \partial_u \nabla H = \frac{1}{3} \nabla \cross \left(H^{-2} \nabla
\cross \left( H^3 {\bf a} \right) \right), \ee The equation of
motion for the field strength turns out to be identically
satisfied.

Equation \p{eqn:Vproj} implies that the Killing spinor satisfies
\be \label{eqn:gammaplusproj}
  \gamma^+ \epsilon = 0.
\ee
Writing out the Killing spinor
equation using the above expressions for the connection and field
strength, and using \p{eqn:gammaplusproj}, one finds that it
reduces to \be
  \partial_\mu \epsilon = 0.
\ee Hence the Killing spinor is constant.
Since the only restriction is
equation \p{eqn:gammaplusproj}, it follows that, in the null case,
as in the timelike case, all supersymmetric solutions preserve at
least half of the supersymmetry.

The above analysis yields the general spacetime that admits a constant
Killing spinor and satisfies the equations for
the field strength. However, the function ${\cal {}F}(u,x)$ is
still completely unrestricted so it is necessary to look at the Einstein
equations for further information. As in the time-like case we can
deduce a lot from the integrability conditions discussed in
appendix B. Working  in the above basis, \p{intcondone} implies
that $E_{\alpha -}=0$ and \p{intcondtwo} give $E_{+i}=E_{ij}=0$.
Hence we just need to impose the $++$ component of the Einstein
equation, which gives \be \label{eqn:{}Feq}
  \nabla^2 {\cal {}F} = 2 H^2 D_u W_{ii} + 2 H W_{(ij)}W_{(ij)} +
\frac{2}{3} H
  W_{[ij]} W_{[ij]},
\ee where \be
  D_u \equiv \partial_u - a_i \nabla_i,
\ee and \be
  W_{ij} = D_u H \delta_{ij} - H \nabla_j a_i.
\ee

The solution is now specified as follows. {}First pick a
harmonic function $H(u,{\bf x})$. Next consider equation
\p{eqn:aeq}. The general solution will be the sum of a particular
integral and the general solution of the homogeneous equation \be
  \nabla \cross \left(H^{-2} \nabla \cross \left( H^3 {\bf a} \right)
  \right) = 0.
\ee This equation can be integrated to give \be \label{eqn:cf}
  \nabla \cross \left( H^3 {\bf a} \right) = H^2 \nabla \phi,
\ee for some function $\phi(u,{\bf x})$. The integrability
condition for this
  is
\be
  0 = \nabla \cdot \left( H^2 \nabla \phi \right) = H \nabla^2 (H\phi),
\ee and hence \be
  \phi = KH^{-1}
\ee for some harmonic function $K(u,{\bf x})$. Next, equation
\p{eqn:cf} can be integrated to determine $H^3 {\bf a}$ up to a
gradient. This gradient arises from the gauge freedom $v
\rightarrow v + g(u,{\bf x})$ and therefore can be removed (see the
next subsection).
So the general solution to equation \p{eqn:aeq} involves a single
additional harmonic function $K$. {}Finally, equation
\p{eqn:{}Feq} can be solved to determine ${\cal {}F}$ up to
another arbitrary harmonic function ${\cal {}F}_0$. Hence the
general solution involves three arbitrary $u$-dependent harmonic
functions $H$,
$K$ and ${\cal {}F}_0$.

Recall that a spacetime is said to be a {\it plane-fronted wave}
if it can be foliated by a family of hypersurfaces $u={\rm
constant}$ such that $du$ is null, geodesic and free of expansion,
rotation and shear. This is indeed the case for our solution. In
other words the general null supersymmetric solution is always a
plane-fronted wave (even though it can be static in special cases,
such as the magnetic string). A plane-fronted wave is said to be a
{\it plane-fronted wave with parallel rays} (pp-wave) if $du$ is
also covariantly constant. {}For our solution, this occurs if, and
only if, $H=H(u)$. {}For such a wave, it can be seen from the
definition of $u$ (equation \p{eqn:udef}) that one can take $H
\equiv 1$ without loss of generality (see next subsection) so this
solution is specified
by just two harmonic functions. It is interesting to note that in
four dimensions, the null supersymmetric solutions are pp-waves
but in five dimensions they are more general plane-fronted waves.

\subsection{Changes of coordinate}

Note that there is some unfixed gauge freedom in the solution. These
correspond to co-ordinate transformations that leave the form of
the metric and the field strength invariant. First,
equation \p{eqn:udef} does not define $u$ uniquely. One could
instead work with $u'=u'(u)$ and $H'=H du'/du$. This also affects the
definition of ${\bf x}$: ${\bf x}' = {\bf x} du/du'$.
Next, in defining $v$ by $V=\partial/\partial v$, there is
freedom to specify the surface
$v=0$. This corresponds to the coordinate transformation
\be
\label{eqn:veq}
  v = v' + g(u,x),
\ee which has the effect of replacing ${\bf a}$ and ${\cal {}F}$
in the solution by \be
  {\bf a'} = {\bf a} - H^{-3} \nabla g, \qquad {\cal {}F}' = {\cal {}F} +
  2 \partial_u g - 2 {\bf a} \cdot \nabla g + H^{-3} \left( \nabla g
\right)^2.
\ee
This gauge-freedom can be used to impose a gauge
condition such as \be
  \nabla \cdot {\bf a}=0 \ ,
\ee or \be
  W_{ii}=0.
\ee
Finally, it is clear from
the definition of the coordinates ${\bf x}$ that there is a gauge
freedom ${\bf x} \rightarrow {\bf x} - {\bf v}(u)$ for an arbitrary
$u$-dependent vector ${\bf v}(u)$. These three gauge transformations
will be used repeatedly to simplify solutions.

There are other coordinate transformations that do change the form of
the
solution but are also useful. The coordinates $x^i$ defined above are
not
arbitrary Cartesian coordinates but are related to the $2$-forms
$X^{(i)}$ by equation
\p{eqn:xidef}.
Note that these coordinates are defined in terms of covariantly
constant $2$-forms in exactly the same way as the Cartesian
coordinates on the transverse two-space of a four dimensional
pp-wave \cite{ehlers:62}.
Any other Cartesian coordinate system on $\bR^3$ will
be related to these coordinates by a $u$-dependent rotation and
translation:
\be \label{eqn:roteq}
  {\bf x} = {\bf O}(u) \cdot {\bf x'} + {\bf v}(u),
\ee where ${\bf O}(u)$ is an orthogonal matrix, which we choose to
have determinant +1. The effect of such a transformation is to
rotate the  $2$-forms: $X^{(i)}\to {\bf O}^{ij}X^{(j)}$.
The derivative $D_u$ is invariant under such a transformation. However,
the general form of the solution does {\it not} transform
covariantly -- the solution as given above is valid only in terms
of the preferred coordinates $x^i$. In particular, the above
transformation affect ${\bf a}$ differently in the metric and
field strength. The new metric will be of the same form of as the
old one but with ${\bf x}$ replaced by ${\bf x}'$ and ${\bf a}$
replaced by
\be
  {\bf a'} = {\bf O}^{-1} \left( {\bf a} + \dot{\bf O} \cdot {\bf x'}
  + \dot{\bf v} \right),
\ee where a dot denotes a $u$-derivative.  The left-hand side of
equation \p{eqn:{}Feq} for the
metric function ${\cal F}$ is unchanged since $\nabla^2=\nabla^{'2}$.
Note
that if one wants to write the right-hand-side in terms of primed
quantities,
one should substitute $W={\bf O}W'{\bf O}^{-1} +H \dot{\bf O}{\bf
O}^{-1}$.
The field strength is given by
\be
  {}F = -\frac{1}{2 \sqrt{3}} \epsilon_{ijk} \left[ H^{-2} \nabla'_j
\left( H^3
a'_k \right) + H \left({\bf O}^{-1} \dot{\bf O} \right)_{jk}
\right] du \wedge {dx'}^i - \frac{\sqrt{3}}{4} \epsilon_{ijk}
\nabla'_k
  H {dx'}^i \wedge {dx'}^j.
\ee
Note that one can reach ${\bf a'}=0$ (and therefore eliminate
$du dx^i$ cross terms from the metric) precisely when ${\bf a}$
satisfies Killing's equation $\nabla_{(i} a_{j)}=0$, i.e, when \be
  {\bf a} = {\bf x} \cross \bomega(u) + {\bf b}(u),
\ee for some vectors $\bomega(u)$ and ${\bf b}(u)$. After changing
coordinate to eliminate ${\bf a}$ from the metric, a $du \wedge
{dx'}^i$ term remains in the field strength so the solution in the
new coordinates is not of the same form as the solution in the
original coordinates.

\subsection{pp-waves}

As mentioned above, pp-waves have $H \equiv 1$. Equation
\p{eqn:aeq} gives \be \nabla \cross {\bf a} = \nabla \phi \ee for
some function $\phi(u,x)$. The integrability condition for this
equation is that $\phi$ must be harmonic \be \nabla ^2 \phi =0.
\ee The solution then takes the form
\bea \label{eqn:ppeq}
  ds^2 &=& {\cal {}F} du^2 + 2dudv - (d{\bf x} + {\bf a} du )^2 \nn
{}F &=&
-\frac{1}{2\sqrt{3}} du \wedge d\phi, \qquad \qquad \nabla \times {\bf
a}
= \nabla \phi. \eea The function ${\cal {}F}$ is given by solving
\p{eqn:{}Feq}.\footnote{
Some similar ten dimensional pp-wave solutions were
given in \cite{bergshoeff:93}.}
Without loss of generality, we
impose the gauge condition $ \nabla \cdot {\bf  a}=0$, so that
\p{eqn:{}Feq} becomes \be \label{eqn:{}Ffeq}
  \nabla ^2 {\cal {}F} =  2  \nabla_{(i} a_{j)}\nabla_{(i} a_{j)}
    + \frac{1}{3}  (\nabla \phi)^2.
\ee

If $\nabla_{(i} a_{j)} = 0$, i.e., if ${\bf a} = {\bf x}
\cross \bomega(u) + {\bf b}(u)$ then one can remove ${\bf a}$ from
the metric by a coordinate transformation as described above, and
$\nabla \phi = -2\bomega(u)$ is independent of ${\bf x}$. The
orthogonal matrix occuring in the coordinate transformation must obey
\be
\label{eqn:Odot}
  \dot{O}_{ij}(u) = - \epsilon_{ikl} O_{jk}(u) \omega_l (u).
\ee
In the new coordinates, the solution is
\bea
\label{eqn:ppleq}
  ds^2 &=& {\cal {}F} du^2 + 2dudv -  d{\bf x'}^2 \nn
{}F &=&
  \frac{1}{\sqrt{3}} {\omega'}_i(u) du \wedge d{x'}^i,
  \qquad \qquad {\nabla'}^2 {\cal {}F} = \frac{4}{3}  {\bomega'}^2.
\eea
where $\bomega'(u) = {\bf O^{-1}}(u) \bomega(u)$. Note that
$\dot{\bomega'} = {\bf O^{-1}}(u) \dot{\bomega}$ using
\p{eqn:Odot}. Hence $\bomega'$ is independent of $u$ if, and only
if, $\bomega'$ is (in this case, ${\bf O}$ can be taken as a rotation
about an axis parallel to $\bomega$, giving $\bomega' = \bomega$).
The maximally supersymmetric plane wave solution arising
from a Penrose limit (see below) is of this type.

Another special case is that in which ${}F=0$, so that the space
is a solution of pure gravity, and the Killing spinors are
covariantly constant. This case was analysed in
\cite{Figueroa-O'Farrill:1999tx}, where it was shown that the
holonomy must be in $\bR^3 \subset SO(4,1)$. Setting ${}F=0$ gives
$\nabla \cross {\bf a} = 0$, so ${\bf a}$ is a gradient and can
therefore be set to zero by $v = v' + g(u,{\bf x})$. The solution
is then \be
  ds^2 = {\cal {}F} du^2 + 2dudv -  d{\bf x}^2, \qquad \nabla^2 {\cal
{}F}
  = 0.
\ee The solution given in \cite{Figueroa-O'Farrill:1999tx} (modulo a
typo)
is related to this by a Euclidean transformation.

In section \ref{subsec:gh}, we saw how the timelike class of minimal
$N=2$, $D=4$ supergravity (with action \p{eqn:4daction})
can be oxidized to give a subset of our
timelike class of minimal $N=1$, $D=5$ supergravity. We can now do the
same for the null class given in \cite{Tod}; this was done in
\cite{meessen2} for the special case of the maximally supersymmetric
plane wave solution.
Consider a pp-wave with ${\bf a} = {\bf x}
\cross \bomega (u)$ and consider a coordinate transformation ${\bf x}
= {\bf O}(u) {\bf x'}$ with
\be
  \dot{O}_{ij}(u) = -\frac{4}{3} \epsilon_{ikl} O_{kj}(u) \omega_l(u).
\ee
Note that this is {\it not} the same coordinate transformation as used
to eliminate ${\bf a}$ from the metric.
In the new coordinates, the solution takes the form
\bea
  ds^2 &=& {\cal {}F} du^2 + 2dudv - \left(d{\bf x'} - \frac{1}{3} {\bf
x'} \cross \bomega' (u) du \right)^2, \nn
F &=& \frac{1}{\sqrt{3}}
\omega'_i du \wedge d{x^i}',
\eea
where $\bomega'(u) = {\bf O^{-1}}(u) \bomega (u) $. Now take $\bomega'
= (\omega'_1(u),0,\omega'_3(u))$ and
let $v = v'-\frac{1}{3} y' (\omega'_3(u) x' -\omega'_1 (u) z')$. The
solution takes the form
\ba
  ds^2 &=& {\cal {}F}' du^2 + 2dudv' - {dx'}^2 - {dz'}^2 - \left(dy' +
  \frac{2}{3} (\omega'_3 x' - \omega'_1 z') du \right)^2, \nn
  \qquad F &=& \frac{1}{\sqrt{3}} \omega'_i(u) du \wedge d{x^i}',
\ea
where
\be
\label{eqn:nullred}
  {\cal {}F}' = {\cal {}F} + \frac{1}{3} (\omega'_3 x' - \omega'_1
  z')^2  - \frac{1}{9}
{\bomega'}^2 {y'}^2 - \frac{2}{3} y' (\dot{\omega}'_3 x' -
  \dot{\omega}'_1 z').
\ee
It is always possible to choose ${\cal {}F}$ obeying \p{eqn:{}Feq} such
that
${\cal {}F}'$ is independent of $y'$. The solution can
then be KK reduced on the Killing vector field $\partial/\partial
y'$. In the language of section \ref{subsec:gh}, we have ${\cal A} =
\frac{2}{3} (\omega'_3 (u) x' - \omega'_1 (u) z') du$
and it can be checked that the consistency conditions for the
  reduction are obeyed. Explicitly, the four dimensional solution is
\bea
\label{eqn:4dnull}
  ds^2 &=& {\cal {}F}' du^2 + 2dudv' - {dx'}^2 - {dz'}^2, \nn
  F' &=& \frac{1}{\sqrt{3}} du \wedge \left( \omega'_1 dx' +
  \omega'_3 dz'\right),
\eea
where ${\cal {}F}'(u,x',z')$, must obey (using \p{eqn:nullred} and
\p{eqn:{}Feq})
\be
  \left( \frac{\partial^2}{\partial {x'}^2} +
  \frac{\partial^2}{\partial {z'}^2} \right)
  {\cal {}F}' = \frac{16}{9} {\bomega'}(u)^2.
\ee

\subsection{Black String, Penrose limits and Plane Wave}

Probaby the best known solutions in the null class are the static
black string solutions \cite{Gibbons:1994vm}.
These have ${\cal F}={\bf a}=0$ and
$H=H({\bf x})$. A single black string is obtained by choosing the
harmonic function $H$ to have a single centre:
\bea
  ds^2&=& H^{-1}\left(2 du dv \right)-H^2 (d{\bf x})^2 \nn
F&=&-\frac{{\sqrt 3}}{4}\epsilon_{ijk}\nabla_k H dx^i\wedge dx^j,\qquad
\qquad
H=1+\frac{R}{2r}
\eea
The tension and magnetic charge per unit length of the string
are both proportional to $R$.

The near horizon limit of this solution is obtained  by taking $r\to 0$
and
hence dropping the 1 in the harmonic function. One then obtains
$AdS_3\times S^2$, which is maximally supersymmetric.
This can be written in global coordinates as
\bea\label{tween}
ds^2&=&R^2[\cosh^2\rho dt^2-d\rho^2-\sinh^2\rho
d\psi^2-\frac{1}{4}(d\theta^2+
\cos^2\theta d\phi^2)]\nn
{}F&=&\frac{ {\sqrt 3}R}{4}\cos\theta d\theta\wedge d\phi
\eea
where $0\le \psi, \phi<2\pi$ and $0\le\theta\le\pi$.

The maximally supersymmetric plane-wave solution presented in
\cite{Meesn} can be obtained as a Penrose limit. Explicitly if we
introduce the new coordinates
\bea
u&=&t+\frac{\phi}{2},\qquad
v=\frac{R^2}{2}(t-\frac{\phi}{2}),\nn
\rho &=&\frac{r}{R},\qquad\qquad
\theta =-\frac{2 z}{R}
\eea
and then take the limit $R\to \infty$ we get
\bea
ds^2&=&2dudv +(z^2+\frac{r^2}{4})du^2
-(dr^2+r^2d\psi^2 +dz^2)\nn
{}F&=&\frac{\sqrt 3}{2}du\wedge dz
\eea
Note that this solution is of the form \p{eqn:ppleq} with
$\bomega=(0,0,3/2)$.
Note also that there is a similar Penrose limit of $AdS_2\times S^3$,
the near horizon
limit of the electric black hole solution, that
gives the same maximal supersymmetric plane-wave solution.

\sect{Maximally supersymmetric solutions}

\label{sec:maximal}

\subsection{Introduction}

The above results for the timelike and null cases
show that Killing spinors always come in pairs
(of Dirac spinors) obeying the same projection \p{eqn:gamma0proj}.
Therefore the solutions all preserve at least half of the
supersymmetry. It is natural to ask which solutions preserve more than
half of the supersymmetry.

A pair spans a two dimensional subspace of spinors. Every spinor in
this subspace gives rise to the same function $f$ and Killing vector
$V$. Let $\epsilon_1$ and $\epsilon_2$ be a pair. If there is an extra
linearly independent Killing spinor $\epsilon_3$ then it must also have
a partner
$\epsilon_4$. Let $f'$, $V'$ denote the function and Killing vector
that arises from this second pair.

Consider first the case in which $\epsilon_4$ is not
linearly independent: $\epsilon_4 = \alpha \epsilon_1 + \beta
\epsilon_2 + \gamma \epsilon_3$, for some functions
$\alpha,\beta,\gamma$.
Since each $\epsilon_i$ is Killing and, by assumption,
$\epsilon_1$, $\epsilon_2$ and $\epsilon_3$ are linearly independent,
we conclude that $\alpha,\beta,\gamma$ are actually constants.
In addition $\epsilon' \equiv
\alpha \epsilon_1 + \beta \epsilon_2$ is a Killing spinor obeying the
same projection as $\epsilon_4$, i.e., it forms a pair with
$\epsilon_3$. It follows that $\epsilon'$ gives rise to the function
$f'$ and Killing vector $V'$.
But, being a linear combination of $\epsilon_1$ and
$\epsilon_2$, $\epsilon'$ must give rise to the Killing vector
$V$. Hence we must have $V = V'$ and similarly $f=f'$ (at least up to a
positive constant of proportionality). However, this implies that
$\epsilon_4$
obeys precisely the same projection (equation \p{eqn:Vproj})
as $\epsilon_1$ and $\epsilon_2$, which contradicts the linear
independence of $\epsilon_3$. Hence $\epsilon_4$ must be linearly
independent of $\epsilon_1$, $\epsilon_2$ and $\epsilon_3$. So if a
solution preserves more than $1/2$ supersymmetry then it must preserve
all supersymmetry.

The goal of this section is to identify those solutions
preserving all supersymmetry. This can be done by examining the
integrability conditions. If there are four independent
Killing spinors then it
is easy to argue that there must exist an open set ${\cal U}$ in which
these
Killing spinors are pointwise linearly independent and hence form a
basis for all spinors. The integrability conditions are algebraic and
must therefore hold for an arbitrary spinor in ${\cal U}$. This yields
an identity of the form
\be
  X_{\alpha \beta \gamma} \gamma^\gamma + Y_{\alpha \beta \gamma
  \delta} \gamma^{\gamma \delta} = 0,
\ee
where $X$ and $Y$ are tensors formed from the field strength and
Riemann tensor. By analytic continuation, this must hold everywhere,
not just in ${\cal U}$.
This expression can only be valid if $X$ and $Y$
vanish separately. Our strategy will be to examine what further
restrictions this gives for the solutions classified above. Rather than
computing $X$ and $Y$ directly from the Riemann tensor and field
strength, it turns out to be simpler to write out the Killing spinor
equation in components and rederive the integrability conditions
component by component, which will clearly give identical results.

\subsection{Maximal null supersymmetry}

In the above analysis, we showed that the null solutions admit Killing
spinors obeying the projection $\gamma^+ \epsilon = 0$. We now want to
find the maximally supersymmetric solutions and must therefore relax
this condition. Any spinor $\epsilon$ can be written as
\be
  \epsilon = \epsilon_+ + \epsilon_-,
\ee
where $\epsilon_+=(1/2)\gamma^+\gamma^-\epsilon$ and
$\epsilon_-=(1/2)\gamma^-\gamma^+\epsilon$.
Substituting the known form
of the null solution into the Killing spinor equation yields the
following components:
\be
\label{eqn:nullv+}
  \partial_v \epsilon_+ + \frac{1}{2} H^{-2} \nabla_i H \gamma^i
  \gamma^+ \epsilon_- = 0,
\ee
\be
\label{eqn:nullv-}
  \partial_v \epsilon_- = 0,
\ee
\be
\label{eqn:nulli+}
  \nabla_i \epsilon_+ - \frac{1}{2} H \left[\frac{1}{3} H
  \nabla_{[i} a_{j]} + W_{(ij)} \right] \gamma^j \gamma ^+ \epsilon_-
  - \frac{1}{6} H^2 \epsilon_{ijk} \nabla_j a_k \gamma^+ \epsilon_- =
  0,
\ee
\be
\label{eqn:nulli-}
  \nabla_i \epsilon_- - H^{-1} \epsilon_{ijk} \nabla_j H \gamma^k
  \epsilon_- = 0,
\ee
\be
\label{eqn:nullu+}
  \left( \partial_u - a^i \nabla_i \right) \epsilon_+ - \frac{1}{4}
\nabla_i \left({\cal F}H^{-1} \right) \gamma^i \gamma^+ \epsilon_- = 0,
\ee
\be
\label{eqn:nullu-}
  \left( \partial_u - a^i \nabla_i \right) \epsilon_- + \frac{1}{3}
  \epsilon_{ijk} \nabla_j a_k \gamma^i \epsilon_- = 0.
\ee
Consider the integrability condition for \p{eqn:nullv+} and
\p{eqn:nulli+}. Using \p{eqn:nullv-} and \p{eqn:nulli-}, this
reduces to
\be
  \left[ H^{-2} \nabla_i \nabla_j H - 3 H^{-3} \nabla_i H
  \nabla_j H + H^{-3} \left(\nabla H \right)^2 \delta_{ij} \right]
  \gamma^j \gamma^+ \epsilon_- = 0.
\ee
We now apply the argument outlined above: $\epsilon_-$ can be replaced
by $\epsilon$ in this expression, and we are demanding that there
exist eight independent solutions. This can only be true if the
expression within square brackets vanishes, which is equivalent to
\be
  \nabla_i \nabla_j H^{-2} = 2 \left (\nabla H^{-1} \right)^2
  \delta_{ij}.
\ee
This equation is easy to solve (e.g. first consider the components
with $i \ne j$) and has solutions $H=H(u)$ or $H^{-2} = f(u)^{-2}
\left({\bf x} + {\bf b}(u) \right)^2$. In the latter case, one can
exploit the unfixed coordinate freedom ${\bf x} \rightarrow {\bf x} -
{\bf c}(u)$ to set ${\bf b} = 0$. In the former case, it was argued
above that one can change coordinates so that $H \equiv 1$. Hence
\be
  H \equiv 1, \qquad {\rm or} \qquad H = \frac{f(u)}{r},
\ee
where $r \equiv \sqrt{{\bf x}^2}$.

Consider first the case $H=f(u)/r$. The integrability condition of
\p{eqn:nullu+} and \p{eqn:nullv+} implies (using \p{eqn:nullv-}
and \p{eqn:nullu-})
\be
\label{eqn:ads1}
  \partial_u \left(H^{-2} \nabla_i H \right) - \frac{H^{-5}}{3}
  \nabla_j H \left[ \nabla_i \left(H^3 a_j \right) - \nabla_j \left(
  H^3 a_i \right) \right] = 0,
\ee
and
\be
\label{eqn:ads2}
  \nabla H \cdot \nabla \cross {\bf a} = 0.
\ee
Multiplying the first equation by $\nabla_i H$ leads immediately to
$\partial_u f = 0$ and hence
\be
  H = \frac{R}{2r},
\ee
for some constant $R$, and a factor of 2 is inserted for convenience.
Equations \p{eqn:ads1} and \p{eqn:ads2}
then imply
\be
  \nabla \cross \left(H^3 {\bf a} \right) = 0,
\ee
which implies that ${\bf a}$ can be gauged away by a transformation of
the form $v \rightarrow v - g(u,{\bf x})$. Hence we can assume ${\bf
a} = 0$. Integrability of \p{eqn:nulli+} and \p{eqn:nullu+} then
implies
\be
  \nabla {\cal F} \cross \nabla H = 0,
\ee
so ${\cal F} = {\cal F}(u,r)$. Equation \p{eqn:{}Feq} implies that
${\cal F}$ is harmonic, hence
\be
  {\cal F} = \frac{f_1(u)}{r} + f_2(u).
\ee
Finally, by considering a coordinate transformation of the form $v =
v' + P_1(u)/r' + P_2(u)$, $r = r'P_3(u)$, $u = P_4(u')$ for appropriate
choices of the functions $P_i$ one can bring the solution to the form
\be
  ds^2 = 4\frac{r'}{R} du' dv' - \frac{R^2}{4} \frac{{dr'}^2}{{r'}^2} -
\frac{R^2}{4}
  d\Omega_2^2, \qquad F = \frac{\sqrt{3}}{4} R {\rm vol}(S^2),
\ee
which is clearly the maximally supersymmetric $AdS_3 \times S^2$
solution \p{tween}. If we drop the primes on the coordinates then this
solution
is simply given by setting ${\cal F} = {\bf a} = 0$, $H=R/2r$ and,
continuing to use the frame \p{milo}, the
Killing spinors are easily found to be
\be
  \epsilon_- = \hat{x}^i \gamma^i \eta_-,
\ee
\be
  \epsilon_+ = \eta_+ + \frac{v}{R} \gamma^+ \eta_-,
\ee
where $\hat{\bf x} = {\bf x}/r$ and $\eta_{\pm}$ are arbitrary
constant spinors obeying $\gamma^{\pm} \eta_{\pm} = 0$.

Now consider the case $H \equiv 1$. Integrability of \p{eqn:nulli-}
and \p{eqn:nullu-} gives
\be
  \nabla \cross {\bf a} = - 2 \bomega (u),
\ee
where $\bomega(u)$ is an arbitrary $u$-dependent vector and hence
\be
  {\bf a} = {\bf x} \cross \bomega(u),
\ee
where an arbitrary gradient can be removed by a shift in
$v$. Integrability of \p{eqn:nulli+} and \p{eqn:nullu+} then
gives
\be
  \bomega = {\rm constant},
\ee
and
\be
  \nabla_i \nabla_j {\cal F} = \frac{2}{9} \left(\delta_{ij} \bomega^2
+ 3 \omega_i \omega_j \right),
\ee
with solution
\be
  {\cal F} = \frac{1}{9} \left( \bomega^2 {\bf x}^2 + 3 \left(\bomega
  \cdot {\bf x} \right)^2 \right) + \balpha(u) \cdot {\bf x} + \beta(u).
\ee
The arbitrary functions $\balpha(u)$ and $\beta(u)$ can be removed by a
combined transformation of the form ${\bf x} \rightarrow {\bf x} -
\bgamma(u)$ and $v \rightarrow v - \blambda(u) \cdot {\bf x} -
\delta(u)$. Hence the final form of the solution is
\ba
  ds^2 &=& \frac{1}{9} \left( \bomega^2 {\bf x}^2 + 3 \left(\bomega
  \cdot {\bf x} \right)^2 \right) du^2 + 2 du dv - \left(d{\bf x} +
  {\bf x} \cross \bomega du \right)^2, \nonumber \\
  F &=& \frac{1}{\sqrt{3}} \omega_i du \wedge dx^i,
\ea
where $\bomega$ is an arbitrary constant $3$-vector. It is now easy to
solve for the Killing spinors:
\be
  \epsilon_- = \left[ \cos \left( 2 \omega u / 3 \right) +
  \hat{\omega}_i \gamma^i \sin \left(2 \omega u /3 \right) \right]
\eta_-,
\ee
\be
  \epsilon_+ = \eta_+ + \frac{1}{6} \left({\bf x} \wedge \bomega
  \right)_i \gamma^i \gamma^+ \epsilon_- - \frac{1}{3} {\bf x} \cdot
  \bomega \gamma^+ \epsilon_-,
\ee
where $\omega = |\bomega|$, $\hat{\bomega} = \bomega/\omega$, and
$\eta_{\pm}$ are arbitrary constant spinors obeying the projections
$\gamma^{\pm} \eta_{\pm} = 0$. In the coordinate system of equation
\p{eqn:ppleq}, the solution is (using $\bomega' = \bomega$)
\be
  ds^2 = \frac{1}{9} \left( \bomega^2 {\bf x'}^2 + 3 \left(\bomega
  \cdot {\bf x'} \right)^2 \right) du^2 + 2 du dv - d{\bf x'}^2, \qquad
  F = \frac{1}{\sqrt{3}} \omega_i du \wedge d{x'}^i.
\ee

We thus conclude that the only maximally supersymmetric solutions in
the null class are $AdS_3 \times S^2$ and the maximally supersymmetric
plane wave. It turns out that both of these solutions also belong to
the timelike class. In other words, some of the Killing spinors
correspond to a null Killing vector but others correspond to a
timelike Killing vector. This is easy to see using the explicit
expressions for the Killing spinors given above. For the plane wave,
take a Killing spinor with $\eta_+ = 0$ and imagine repeating the
analysis of this paper with this spinor as the fiducial spinor. One
then obtains
\be
  f = -\frac{\sqrt{2}}{3} {\bf x} \cdot \bomega \eta_-^\dagger \eta_-,
\ee
which is clearly non-zero (although we need to restrict to ${\bf x}
\cdot \bomega < 0$ for $f>0$). Similarly, for the $AdS_3 \times S^2$
solution one obtains from the general Killing spinor
\be
  f = -\frac{1}{\sqrt{2}} \hat{x}^i \Re \left( \eta_-^\dagger \gamma^i
  \gamma^- \eta_+ \right),
\ee
where $\Re$ denotes the real part,
which is also non-zero in general. These solutions can be cast into
the timelike form of section \ref{sec:timelike} as follows.

For the plane wave, a Killing spinor
with $\eta_+ = 0$ gives the Killing vector
\be
  V = \frac{\partial}{\partial u} - \frac{4}{3} a_i
  \frac{\partial}{\partial x^i},
\ee
where we have normalized so that $\eta_-^\dagger \eta_- =
\sqrt{2}$. We want to write the solution in the timelike form, so we
need to choose new coordinates $(t,y^m)$ such that $V =
\partial/\partial t$. A natural guess is to choose new coordinate
$(t,x^5,{\bf x}')$ where $t = u$, $x^5 = v$ and
\be
  \frac{\partial x^i}{\partial t} = - \frac{4}{3} a_i.
\ee
A solution is to take ${\bf x} = {\bf O}(t) {\bf x}'$ where ${\bf
O}(t) = \exp ( {\bf A} t )$ is orthogonal and the antisymmetric matrix
${\bf A}$ is given by
\be
  A_{ij} = -\frac{4}{3} \epsilon_{ijk} \omega_k.
\ee
Note that the same coordinate transformation was used above in the
dimensional reduction of the null class. In these new
coordinates, the solution takes the form
\bea
\label{eqn:nullitya}
v ds^2 &=& f^2 \left (dt + \omega \right)^2 - f^{-1} \left[ f^{-1}
\left(dx^5 + \bchi\cdot d{\bf x'} \right)^2 + f d{\bf
  x'}^2 \right], \nonumber \\
  F &=& \frac{1}{\sqrt{3}} \omega_i dt \wedge {dx'}^i \ ,
\eea
where
\be
  f = - \frac{2}{3} {\bf x'} \cdot \bomega, \qquad \bchi =
\frac{1}{3}{\bf x'}
\cross \bomega, \qquad
  \omega = f^{-2} \left(dx^5 + \bchi \cdot d{\bf x'}
  \right).
\ee
The solution is now written in the timelike form. The base space is a
Gibbons-Hawking space with a linear harmonic function $H=f$,
corresponding to a constant density planar distribution of Taub-NUT
instantons. In the language of our general analysis of solutions with
a Gibbons-Hawking base space, this solution has $K=1$ and
$L=M=\bomega=0$. Note that $K/H$ is not constant, so this solution has
$G^+ \ne 0$.

For $AdS_3 \times S^2$, the Killing vector $V$ constructed from the
general Killing spinor turns out to be rather complicated, but it is
possible to proceed by trial and error. First, note that we can rotate
$S^2$ and normalize the spinor so that $f = \cos \theta$. Now the
solution can be massaged into timelike form by writing $AdS_3$ in a
form with ${\cal F} = R/2r$ (as discussed above):
\bea
  ds^2 &=& du^2 + 4\frac{r}{R} du dv - \frac{R^2}{4r^2} dr^2 -
\frac{R^2}{4}
  \left(d\theta^2 + \sin^2 \theta d\phi^2 \right),\nn
F &=&
\frac{\sqrt{3}}{4} R \sin \theta d\theta \wedge d\phi.
\eea
Letting $u=t$, $v=x^5$, $\phi = \phi' - 2t/R$, the solution can be
written
\ba\label{ghdip}
  ds^2 &=& f^2 \left(dt + \omega \right)^2 - f^{-1}
  \left[H \left( dr^2 + r^2 d\theta^2
  + r^2 \sin^2 \theta d{\phi'}^2 \right) + H^{-1}
  \left(dx^5 + \frac{R^2 \sin^2 \theta}{4r} d\phi' \right)^2 \right],
  \nonumber \\
  F &=& \frac{\sqrt{3}}{2} \sin \theta \left( dt \wedge d\theta +
  \frac{R}{2} d\theta \wedge d\phi' \right),
\ea
where
\be
  f = \cos \theta, \qquad \omega = \frac{2r}{R \cos^2 \theta}
  \left(dx^5 + \frac{R^2 \sin^2 \theta}{4r} d\phi' \right), \qquad H =
  \frac{R^2 \cos \theta}{4r^2}.
\ee
The base space is a Gibbons-Hawking solution with a dipole source for
$H$. $K$ has a monopole source: $K = R/2r$ and
$L=M=\bomega=0$. Once again, $K/H$ is not constant so this
solution has $G^+ \ne 0$.

It is noteworthy that both of these regular
maximally supersymmetric solutions can be cast in the time-like form
using singular hyper-K\"ahler base spaces.
Note also that for both solutions, the
null Killing vector field $\partial/\partial v$ coincides with the
triholomorphic Killing vector field $\partial/\partial x^5$. Both
solutions
are special cases of the class of timelike solutions given in section
\ref{subsec:gh} with $L=M=\bomega=0$ (so
$f^{-1} = K^2/H$ and $\omega_5 = (K/H) f^{-1}$) for which
$\partial/\partial x^5$ is null.

It was remarked in \cite{meessen2} that $AdS_3 \times S^2$ can be
obtained as a limit of the maximally supersymmetric near horizon
geometry of the BMPV black hole discussed in
\cite{gauntlett:99}. It is interesting to see how this works in
our framework. The near horizon geometry of BMPV has a flat base
space and $G^+ = 0$. In Gibbons-Hawking form, $H$ has a monopole
source and $f^{-1}$ and $\omega_5$ are proportional to $H$. More
generally, consider a  solution with Gibbons-Hawking base,
$f^{-1} = H$ and $\omega_5 = cH$ for some constant $c$:
\ba
  ds^2 &=& H^{-2} dt^2 + 2cH^{-1} dt \sigma - (1-c^2) \sigma^2 - H^2
  d{\bf x}^2, \qquad \sigma = dx^5 + \chi_i dx^i,\nonumber \\
  F &=& \frac{\sqrt{3}}{2} \left(-dt \wedge d(H^{-1}) + \frac{c}{2}
\epsilon_{ijk} \nabla_k H dx^i \wedge dx^j \right),
\ea
{}For $c^2 < 1$ one can KK reduce on $\partial/\partial x^5$
as described in section \ref{subsec:gh} (after rescaling co-ordinates).
If $c^2=1$ then $\partial/\partial x^5$ is
null and this reduction is no longer possible (although the metric
remains
non-degenerate). However, one can introduce new coordinates
\be
  t' = \frac{t}{\sqrt{1-c^2}}, \qquad {x^5}' = \sqrt{1-c^2} x^5,
\ee
and then take the limit $c^2 \rightarrow 1$ with $t'$ and ${x^5}'$ held
fixed. This is not the same as setting $c^2 = 1$ in the original
solution. For $c \rightarrow 1$ we get
\be
  ds^2 = H^{-1} \left( -H {d{x^5}'}^2 + 2 dt' d{x^5}' \right) - H^2
  d{\bf x}^2, \qquad F = \frac{\sqrt{3}}{4} \epsilon_{ijk} \nabla_k H
  dx^i \wedge dx^j,
\ee
which is a solution belonging to the null class with $u={x^5}'$,
$v=t'$, ${\cal F} = -H$ and ${\bf a} = 0$.\footnote{The observant
reader will notice a discrepancy in the sign of the field strength
between this equation and \p{eqn:nullFsol}. This arises because the
null solutions have no preferred orientation. The sign can be fixed
by ${\bf x} \rightarrow -{\bf x}$.} This limit
must therefore involve a boost.
If $H$ has a monopole source then the above procedure takes the near
horizon geometry of BMPV to $AdS_3 \times S^2$ written in null
form.

\subsection{Maximal Timelike Supersymmetry}

We now determine the maximally supersymmetric
solutions which are associated with
a timelike killing vector by analysing
the Killing spinor equations
in the timelike background described in section 3.
We decompose the Killing spinor as
$\e= \e^- +f^{1 \over 2} \e^+$
where $\g^0 \e^{\pm} = \pm \e^\pm$. Then the Killing spinor equations
may be written as
\be
\label{eqn:kspxa}
\pd{t} \e^+ =- \g^i \nabla_i f \e^-
\ee

\be
\label{eqn:kspxb}
\pd{t} \e^- = {f^2 \over 6} \g^{ij} G^+{}_{ij} \e^-
\ee

\be
\label{eqn:kspxc}
\nabla_i \e^+ +\omega_i \g^j \nabla_j f \e^- -{1 \over 3}G^+{}_{ij}
\g^j \e^- -G^-{}_{ij} \g^j \e^-=0
\ee
and

\be
\label{eqn:kspxd}
\nabla_i \e^- -{f^2 \over 6} \omega_i G^+{}_{jk} \g^{jk} \e^- +{1 \over
2}f^{-1} \delta_{ij}
\nabla_k f \g^{jk} \e^- +{1 \over 2} f^{-1} \nabla_i f \e^-=0
\ee
In the above equations, all spatial indices are with respect to an
orthonormal basis
with respect to the base space metric $h$, and $\nabla$ is the
covariant derivative
with respect to $h$.

To proceed, we evaluate the integrability conditions of these
equations. First,
for maximal supersymmetry, the integrability condition of
({\ref{eqn:kspxb}}) and
({\ref{eqn:kspxd}}) gives
\be
\label{eqn:integxa}
\nabla_i (f^2 G^+{}_{pq})-f \delta_{ip} \nabla_k f \delta^{k \ell}
G^+{}_{lq}
+f \delta_{iq} \nabla_k f \delta^{k \ell} G^+{}_{lp}+f \nabla_p f
G^+{}_{iq}
-f \nabla_q f G^+{}_{ip}=0
\ee
The integrability condition between ({\ref{eqn:kspxa}}) and
({\ref{eqn:kspxc}})
is
\be
\label{eqn:integxb}
- \nabla_i \nabla_j f +{1 \over 2f} \delta_{ij} \delta^{pq} \nabla_p f
\nabla_q f
+{2 \over 3}f^2 ({1 \over 3} G^+{}_{ip}+ G^-{}_{ip})\delta^{pq}
G^+{}_{q j}=0
\ee
The integrability condition of ({\ref{eqn:kspxd}}) is
\bea
\label{eqn:curvint}
C_{ijpq} &=& f \big[ -{2 \over 3} G^+{}_{ij} G^+{}_{pq} +{1 \over 18}
G^+{}_{mn}G^{+mn} (\delta_{ip}
\delta_{jq} - \delta_{jp} \delta_{iq} + \epsilon_{ijpq}) \big]
\eea
where $C$ is the Weyl tensor, and the integrability condition of
({\ref{eqn:kspxc}}) is
\bea
\label{eqn:finaldif}
- f\nabla_\ell G^-{}_{ij} +  \big(\nabla_i f G^-{}_{j \ell} - \nabla_j
f G^-{}_{i \ell}- \nabla_\ell
f G^-{}_{ij} \big) + \big( \delta_{j \ell} G^-{}_{im}\delta^{mn}
\nabla_n f
-  \delta_{i \ell} G^-{}_{jm}\delta^{mn} \nabla_n f\big)
\nn
+ {1 \over 3}  (\nabla_\ell f G^+{}_{ij} +\nabla_i fG^+{}_{j \ell} -
\nabla_j fG^+{}_{i \ell})
-{1 \over 3} \big( \delta_{j \ell} G^+{}_{im}\delta^{mn} \nabla_n f
-  \delta_{i \ell} G^+{}_{jm}\delta^{mn} \nabla_n f\big) =0
\eea

Let us first investigate the special case $G^+=0$.
{}From ({\ref{eqn:curvint}}) we immediately
conclude that the base space is flat. In fact, the square of
\p{eqn:curvint} relates the square of the Weyl tensor of the base
space to the square of $G^+$ and hence a maximally supersymmetric
solution has flat base space if, and only if, $G^+ = 0$.
Solving ({\ref{eqn:integxb}}) we find that the two possible solutions
are:
\bea
\label{eqn:flatfsol}
f&=&\alpha\,\, or, \\
f&=&{\alpha \over 2}x^m x^m = {\alpha \over 2}r^2
\eea
for $\alpha>0$ constant.

We consider first the case $f=\alpha$. Then ({\ref{eqn:finaldif}})
implies
that $G^-$ is covariantly constant. Now
any anti-self-dual two form can be expanded in terms of the
standard anti-self-dual complex structures $J^{(i)}= {1 \over 4}
d \big[ r^2 \sigma^i_L \big] $ on $\bR^4$
as $G^-=\lambda^i J^{(i)}$ and we deduce that
the $\lambda^i$ are constants; so
\bea
\omega= {\lambda^i r^2 \over 4 \alpha } \sigma^i_L
\eea
The five-dimensional metric is given by
\be
\label{eqn:flatfivea}
ds^2 = \alpha^2 (dt+{\lambda^i r^2 \over 4 \alpha }
\sigma^i_L)^2-\alpha^{-1}
\big[ dr^2 +r^2 d \Omega_3{}^2 \big]
\ee
and is the maximally supersymmetric G\"odel type solution
investigated previously.

Let us now consider the case $f={\alpha \over 2} r^2$. We introduce
a new basis of anti-self-dual two forms
$Q^{(i)}=d \big[r^{-2} \sigma^i_R \big]$.
Then writing $G^- = \lambda^i r^2 Q^{(i)}$ we
find on substituting into ({\ref{eqn:finaldif}})
that the $\lambda^i$ must be constant. Hence
\be
\label{eqn:omegasol}
\omega = {2 \over \alpha r^2} \lambda^i \sigma^i_R
\ee
The five dimensional spacetime geometry is given by
\be
\label{eqn:nhbmpv}
ds^2 = {\alpha^2 \over 4}r^4 (dt + {2 \over \alpha r^2}
\lambda^i \sigma^i_R)^2
-{2 \over \alpha r^2} \big[ dr^2 +r^2 d \Omega_3{}^2 \big]
\ee
This geometry is the near-horizon geometry of the rotating
BMPV five-dimensional black hole which was shown to be maximally
supersymmetric
in \cite{gauntlett:99}. Setting $\lambda^i = 0$ gives $AdS_2 \times
S^3$.

\subsubsection{Maximal supersymmetry with $G^+ \neq 0$}

If $G^+\ne 0$ then the above equations are much more complicated.
Before solving these equations, it is useful to
consider two examples of maximally supersymmetric solutions
with $G^+ \neq 0$. Both examples use a singular hyper-K\"ahler base,
the first a singular version of Eguchi-Hanson and the second negative
mass Taub-NUT.
Surprisingly, both examples are related by co-ordinate transformations
to
time-like solutions built from a flat base space with  $G^+=0$. This
emphasizes the point that $G^+$ is defined with respect to a
particular four dimensional base space, which in turn is defined
by a timelike Killing vector constructed from a Killing spinor.
{}For solutions with maximal supersymmetry it is possible
that the extra Killing spinors give rise to a different timelike Killing
vector and hence a different base space, and in this case
there will then be no simple relation between the old and new $G^+$.

Consider first the singular Eguchi-Hanson Solution
with base metric on $B$ given by
\be\label{singeh}
ds^2 = W^{-1}dr^2+ {r^2 \over 4}((\sigma^1_L)^2+(\sigma^2_L)^2)
+{r^2 \over 4} W (\sigma^3_L)^2
\ee
where $W=1+{b^4 \over r^4}$. Take the solution given by \p{lasty} with
$\delta=\lambda=\gamma=0$:
\bea
f^{-1}&=& \frac{\chi^2}{9b^4 r^2},\qquad
\omega = -{\chi^3 \over 54 b^4 r^4} \sigma^3_L,\nn
G^+ &=& -{\chi \over 4} d (r^{-2} \sigma^3_L),\qquad
G^- = {\chi \over 6r^3}(dr \wedge \sigma^3_L -
{r \over 2} \sigma^1_L \wedge \sigma^2_L).
\eea
It is straightforward to show that all of the
integrability conditions given above are satisfied.
The metric is
\be
  ds^2 = \left(\frac{3b^2}{\chi}\right)^4 r^4dt^2 - \frac{3b^4}{\chi} dt
  \sigma^3_L - \frac{\chi^2}{9 b^4 r^2} \left( 1 + \frac{b^4}{r^4}
  \right)^{-1} dr^2 - \frac{\chi^2}{36 b^4}
  \left[(\sigma^1_L)^2 + (\sigma^2_L)^2 + (\sigma^3_L)^2 \right].
\ee
Now perform a coordinate transformation
\be
  dv = dt + {}F(r) dr, \qquad d \phi' = d\phi + G(r) dr \Rightarrow
{\sigma^3_L}' = \sigma^3_L + G(r) dr,
\ee
with ${}F$ and $G$ chosen so that the coefficients of $dr^2$ and $dr
(\sigma^3_L)'$ vanish. The new metric is
\be
  ds^2 = \left (\frac{3b^3}{\chi} \right)^4 \left( 1 + \frac{r^4}{b^4}
\right) dv^2 - \frac{6 b^2 r}{\chi} dv dr - \frac{\chi^2}{36 b^4}
\left[ (\sigma^1_L)^2 + (\sigma^2_L)^2 + \left({\sigma^3_L}' +
\frac{54 b^8}{\chi^3} dv \right)^2 \right].
\ee
Finally, let
\be
  \phi'' = \phi' + \frac{54 b^8}{\chi^3} v, \qquad v' = \left(
  \frac{3b^3}{\chi} \right)^2 v, \qquad \rho = \frac{\chi r^2}{6 b^4}.
\ee
The metric is now
\be
  \left[ 1 + \left( \frac{6b^2}{\chi} \right)^2 \rho^2 \right] {dv'}^2 -
  2dv'd\rho - \frac{\chi^2}{36 b^4} \left[d\theta ^2 + \sin^2 \theta
  d\psi^2 + \left(d\phi'' + \cos \theta d\psi \right)^2 \right].
\ee
This is clearly $AdS_2 \times S^3$ where the radius of the
$AdS_2$ is given by $\chi/(6b^2)$ and the radius of the $S^3$
by $\chi/(3b^2)$. Note that $r$ corresponds to the {\it global} radial
coordinate $\rho$, with $r=0$ the origin of $AdS_2$. This contrasts with
the description of $AdS_2 \times S^3$ with a flat base space, for
which $r$ is the horospherical ``radial'' coordinate.

Our next example has negative mass Taub-NUT as its base space.
The solution is given by setting
$\gamma=0$, $a=-b<0$ in \p{bad}. Explicitly, the base space metric is
\be\label{negtn}
ds^2 = {(r-b) \over (r+b)} dr^2
+(r^2-b^2)((\sigma^1_R)^2+(\sigma^2_R)^2)
+4b^2 {(r-b) \over (r-b)} (\sigma^3_R)^2
\ee
and the solution is given by
\bea
  f^{-1}&=&{2\chi^2\over
9b(r-b)},\qquad
\omega= {\chi^3 \over 27 b^2} {(r-5b)(r+b) \over (r-b)^2} \sigma^3_R\nn
G^+ &=& \chi d \big[ {(r+b) \over (r-b)} \sigma^3_R \big],\qquad
G^- = -{\chi \over 6b} {(r+b) \over (r-b)^2} \big[2b dr \wedge
\sigma^3_R
+(r^2-b^2) \sigma^1_R \wedge \sigma^2_R \big]\ .
\eea
It is straightforward to show that these expressions satisfy the
integrability conditions.
Surprisingly this solution is just the maximally supersymmetric G\"odel
type solution. To see this we first note that in going from positive
mass
parameter Taub-NUT to negative mass Taub-NUT there is a change
of orientation. Hence it is natural to work with right invariant
one-forms
rather than left-invariant one-forms. This is simply achieved by
interchanging the coordinates $\phi$ and $\psi$. If we do this then the
metric becomes
\bea
ds^2 &=& {81 b^2 \over 4 \chi^4} (r-b)^2 dt^2 -{3 \over 2
\chi}(r-5b)(r+b)
dt \sigma^3_L -{2 \chi^2 \over 9 b(r+b)} dr^2
\nn
&-&{2 \chi^2 \over 9b}(r+b)
((\sigma^1_L)^2+(\sigma^2_L)^2) +{\chi^2 \over 36b^2}(r+b)(r-7b)
(\sigma^3_L)^2
\eea
Let $\phi' = \phi - (3/\chi)^3 b^2 t$ and $r = -b + \rho^2/(8b)$. The
metric becomes
\ba
  ds^2 &=& \left( \frac{9 b^2}{\chi^2} \right)^2 \left[dt -
\left(\frac{\chi}{3} \right)^3 \left(\frac{\rho^2}{16 b^4} \right)
\left(d\phi' + \cos \theta d\psi \right) \right]^2 \nn
  &-& \left(
\frac{9b^2}{\chi^2} \right)^{-1} \left[d\rho^2 + \frac{\rho^2}{4}
\left(d\theta ^2 + \sin^2 \theta d\psi^2 + \left(d\phi' + \cos \theta
  d\psi \right)^2 \right)\right],
\ea
which, after rescaling $t$ and $\rho$, is the generalized G\"odel
solution \p{godsolution} with $\gamma = -3/16 \chi$.

To proceed with finding the maximally supersymmetric timelike solutions
with
$G^+ \neq 0$ it is convenient to prove the

{\bf Proposition.} The hyper-K\"ahler base space $B$ of
the maximally supersymmetric solutions is Gibbons-Hawking.
Moreover, the tri-holomorphic Killing vector is a Killing
vector of the five-dimensional solution.

\noindent {\it Proof.}

Using ({\ref{eqn:integxa}}), ({\ref{eqn:integxb}}) and
({\ref{eqn:finaldif}})
it follows that
\be
\label{eqn:killa}
K^i =f \big(G^{+ij}-3 G^{-ij}\big) \nabla_j f
\ee
satisfies $\nabla_{(i} K_{j)}=0$ and $\cL_K X^{(i)}=0$.
So if $K \neq 0$, as for the negative mass Taub-NUT example
presented in the previous section,
it follows that the base space is Gibbons-Hawking.
Moreover, it is clear that $\cL_K f =0$. In order to show that this
solution falls into the classification of Gibbons-Hawking solutions
presented previously, we also require $\cL_K \omega=0$.
In fact, it suffices to show locally that $\cL_K d \omega=0$.
To do this, we note that  on contracting ({\ref{eqn:integxa}}) with
  $f^2 G^{+pq}$ we find $z^2= f^4 G^+{}_{ij} G^{+ij}$ is constant; $z
\neq 0$.
Then it is straightforward to see from the integrability constraints
that
\be
\label{eqn:wconserva}
f^{-1} K^j \big(G^+{}_{ij}+G^-{}_{ij} \big) = \nabla_i \big(
{z^2 \over 12} f^{-3}- {3 \over 4}f G^-{}_{mn} G^{-mn} \big)
\ee
and hence $d (i_K d \omega)=0$.

It is however also possible that $K=0$, as it is for the singular
Eguchi-Hanson example discussed in the previous section.
To proceed in this case, we note that $K=0$ together with
({\ref{eqn:finaldif}}) implies that $f^2 G^-$ is covariantly constant.
It is then convenient to define
\be
\label{eqn:holokill}
{\hat{K}}_i = f^2 G^-{}_{ij} \nabla^j f\ .
\ee
Note that if ${\hat{K}}$ vanishes, or $f$ is constant, then the base
space must be flat.
Hence we shall consider ${\hat{K}} \neq 0$. It is straightforward to
show that ${\hat{K}}$ is a Killing vector.
Furthermore, without loss of generality we have $f^2 G^- = {z \over 6}
X^{(1)}$.
Next, note that ({\ref{eqn:integxb}})
implies that
\be
\label{eqn:auxident}
{1 \over 2} f^{-1} \nabla^i f \nabla_i f = {z^2 \over 18} f^{-2} +
\alpha
\ee
for constant $\alpha$. Hence we find that
\be
\label{eqn:dholkil}
\nabla_i {\hat{K}}_j = -{ \alpha z \over 6} X^{(1)}_{ij} + {z^2 \over
54}  G^+{}_{ij}
\ee
Furthermore,
when $K=0$ the integrability conditions imply the following
useful identities:
\bea
\label{eqn:potent}
\omega &=& {1 \over f(\alpha f^2+{z^2 \over 18})} {\hat{K}} \ , \qquad \
3 G^- +G^+ =-{3 \over f(\alpha f^2+{z^2 \over 18})}df \wedge {\hat{K}}
\nn
G^+ &=& {3 \over 2} d \big[ {1 \over \alpha f^2+{z^2 \over 18}}
{\hat{K}} \big] \ , \qquad \
G^- =-{1 \over 2 f^2} d \big[ {f^2 \over \alpha f^2+{z^2 \over 18}}
{\hat{K}} \big]
\eea

To proceed we define the following vector fields;
\bea
\label{eqn:coordvects}
S^i &=&{f \over 2(\alpha f^2+{z^2 \over 18})} \nabla^i f \ , \quad
(\sigma^1)^i = {f \over (\alpha f^2+{z^2 \over 18})} (X^1)^{ij}
\nabla_j f
\nn
(\sigma^2)^i &=&  (\alpha f^2+{z^2 \over 18})^{-{1 \over 2}}
(X^2)^{ij} \nabla_j f
\ , \quad  (\sigma^3)^i =  (\alpha f^2+{z^2 \over 18})^{-{1 \over 2}}
(X^3)^{ij} \nabla_j f
\eea
and we note that ${\hat{K}}={z \over 6f}  (\alpha f^2+{z^2 \over 18})
\sigma^1$, so $\omega ={z \over 6 f^2}
\sigma^1$ and $G^- = -{z \over 12 f^2} d(f \sigma^1)$.
In addition, we note that the following constitutes an orthonormal
basis of 1-forms;
\be
\label{eqn:orthobas}
e^1 = \sqrt{f \over 2 (\alpha f^2+{z^2 \over 18})} df \ ,
\quad
e^2 = \sqrt{ (\alpha f^2+{z^2 \over 18}) \over 2f} \sigma^1 \ ,
\quad
e^3 =\sqrt{f \over 2} \sigma^2 \ ,
\quad
e^4 =\sqrt{f \over 2} \sigma^3
\ee
and so the metric on the base space is
\bea
\label{eqn:basemet}
ds^2 = {f \over 2(\alpha f^2+{z^2 \over 18})} df^2 + { (\alpha f^2+{z^2
\over 18}) \over 2f} (\sigma^1)^2
+{f \over 2}\big( (\sigma^2)^2+ (\sigma^3)^2 \big)
\eea
where as a consequence of the integrability conditions the $\sigma^i$
satisfy
\be
\label{eqn:sigmaeq}
d \sigma^1 = \sigma^2 \wedge \sigma^3 \ , \quad \ d \sigma^2 = \alpha
\sigma^3 \wedge \sigma^1 \ , \quad
d  \sigma^3 = \alpha \sigma^1 \wedge \sigma^2 \ .
\ee
To continue, it is useful to introduce some local co-ordinates. In
particular, we find from the integrability
conditions that $[S , {\hat{K}} ]=0$ and so we can introduce local
co-ordinates $y$ and $\phi$ such that
\be
\label{eqn:localcoords}
S = {\partial \over \partial y} \ , \qquad
  {\hat{K}} = {\partial \over \partial \phi}
\ee
and let the remaining two co-ordinates be $\theta$, $\psi$. In
particular, as $S(f)=1$ and ${\hat{K}}(f)=0$, it follows that
$f=y+Q(\theta, \psi)$. Moreover, we note that $i_S \sigma^i=i_
{\hat{K}} \sigma^2 = i_ {\hat{K}} \sigma^3=0$
and $i_ {\hat{K}} \sigma^1 = {z \over 3}$ and therefore we find that
  $\cL_S  \big( (\sigma^2)^2+ (\sigma^3)^2
\big) = \cL_ {\hat{K}}  \big( (\sigma^2)^2+ (\sigma^3)^2
\big)=0$. Hence we can write
\bea
\label{eqn:basemetsimp}
   \big( (\sigma^2)^2+ (\sigma^3)^2 \big) &=& {\cal{H}}(\theta, \psi)^2
  (d \theta^2+ \sin^2 \theta d \psi^2)
   \nn
   \sigma^1 &=&{z \over 3}d \phi + {\cal{P}}(\theta,\psi) d \psi
\eea
where ${\cal{H}}$, ${\cal{P}}$ are constrained by the integrability
conditions.
In particular, requiring that $f^2 G^-$ be covariantly constant
together with the vanishing of the Ricci tensor implies
\be
\label{eqn:constrxa}
{\partial {\cal{P}} \over \partial \theta}  = \pm {\cal{H}}^2 \sin
\theta
\ee
together with
\be
\label{eqn:constrxb}
\Box \log {\cal{H}} = 1- \alpha {\cal{H}}^2
\ee
where $\Box$ denotes the Laplacian defined with respect to the 2-metric
$ds_2{}^2 =d \theta^2+
\sin \theta^2 d \psi^2$.
It is convenient to
write the metric on the unit 2-sphere in terms of complex co-ordinates
$Z$, ${\bar{Z}}$
where $Z= \cot {\theta \over 2} e^{i \psi}$, ${\bar{Z}}= \cot {\theta
\over 2} e^{-i \psi}$;
\be
\label{eqn:unitiisp}
ds_2{}^2 = d \theta^2+ \sin \theta^2 d \psi^2 = {4 \over (1+Z
{\bar{Z}})^2} dZ d {\bar{Z}}
\ee
then ({\ref{eqn:constrxb}}) can be written as $(1+Z {\bar{Z}})^2
{\partial ^2 \over \partial Z
\partial {\bar{Z}}} \log {\cal{H}} = 1-\alpha {\cal{H}}^2$. On setting
  ${\cal{H}}=(1+Z {\bar{Z}}) {\cal{G}}$
we observe that
\be
\label{eqn:liou}
{\partial ^2 \over \partial Z
\partial {\bar{Z}}} \log {\cal{G}}
=-\alpha {\cal{G}}^2 \ .
\ee

There are therefore three cases to consider. In the first, $\alpha =0$
and so
${\cal{G}} = e^{{\cal{F}} + {\bar{\cal{F}}}}$ where ${\cal{F}}(Z)$ is
holomorphic in $Z$.
Hence, by making a holomorphic co-ordinate transformation, we can set
${\cal{G}}=1$ which corresponds to
taking ${\cal{H}}=\sin^{-2} {\theta \over 2}$, ${\cal{P}} = \mp 2
\sin^{-2} {\theta \over 2}$. The base metric
is
\be
\label{eqn:basea}
ds^2 = {9 \over z^2}f df^2 +{z^2 \over 36 f} ({z \over 3} d \phi \mp  2
\sin^{-2} {\theta \over 2} d \psi)^2
+{f \over 2 \sin^4 {\theta \over 2}} (d \theta^2+ \sin^2 \theta d
\psi^2)
\ee
and it is straightforward to show that this metric is Gibbons-Hawking
with tri-holomorphic Killing vector
${\partial \over \partial \phi}$, which clearly preserves $\sigma^1$.

In the second case, $\alpha >0$ and so the general solution to the
Liouville equation
({\ref{eqn:liou}}) is ${\cal{G}}^2 = \alpha^{-1} (1+ {\cal{F}}
{\bar{\cal{F}}})^{-2}
{d {\cal{F}} \over dZ}  {d  {\bar{\cal{F}}}  \over d {\bar{Z}}}$ where
${\cal{F}} (Z)$ is holomorphic in $Z$.
So by making a holomorphic change of co-ordinates we can set
${\cal{H}}={1 \over \sqrt{\alpha}}$
which corresponds to ${\cal{P}} = \mp \alpha^{-1} cos \theta$. The
metric on the base is then
\be
\label{eqn:baseb}
ds^2 = {f \over 2} (\alpha f^2+{z^2 \over 18})^{-1} df^2 +{f \over 2
\alpha}(d \theta^2+ \sin^2 \theta
d \psi^2)+{1 \over 2f} (\alpha f^2+{z^2 \over 18})({z \over 3}d \phi
\mp  \alpha^{-1} cos \theta d \psi)^2
\ee
which is Gibbons-Hawking with tri-holomorphic Killing vector ${\partial
\over \partial \psi}$
which preserves $\sigma^1$.

In the last case, $\alpha<0$, and so on setting $\beta = - \alpha$,
the general solution
to the Liouville equation
({\ref{eqn:liou}}) is ${\cal{G}}^2 = \beta^{-1} ({\cal{F}}+
{\bar{\cal{F}}})^{-2}
{d {\cal{F}} \over dZ}  {d {\bar{{\cal{{F}}}}} \over d {\bar{Z}}}$
where ${\cal{F}} (Z)$ is holomorphic in $Z$.
Hence, by making a holomorphic co-ordinate transformation, we can take
${\cal{H}}= {1 \over \sqrt{\beta} \sin \theta \cos \psi}$. Then the
base metric is given by
\be
\label{eqn:baseca}
ds^2 =  {f \over 2} (- \beta f^2+{z^2 \over 18})^{-1} df^2 +{f \over 2
\beta \sin^2 \theta
\cos^2 \psi }(d \theta^2+ \sin^2 \theta
d \psi^2)+{1 \over 2f} (-\beta f^2+{z^2 \over 18})({z \over 3}d \phi +
{\cal{P}} d \psi)^2
\ee
where ${\partial {\cal{P}} \over \partial \theta} = \pm {1 \over \beta
\sin \theta \cos^2 \psi}$.
On making a change of co-ordinates $d \theta = \sin \theta d \chi$
together with a shift in $\phi$ this metric
can be rewritten as
\be
\label{eqn:basec}
ds^2 =  {f \over 2} (- \beta f^2+{z^2 \over 18})^{-1} df^2 +{f \over 2
\beta \cos^2 \psi }(d \chi^2+
d \psi^2)+{1 \over 2f} (-\beta f^2+{z^2 \over 18})({z \over 3}d \phi
\mp \beta^{-1} \tan \psi d \chi)^2
\ee
which is  Gibbons-Hawking with tri-holomorphic Killing vector
${\partial \over \partial \chi}$
which preserves $\sigma^1$. {\bf{Q.E.D}}.

\subsubsection{Maximally Supersymmetric Gibbons-Hawking Solutions}

We have shown that in all cases the base space corresponding to the
maximally supersymmetric timelike solutions is Gibbons-Hawking,
and moreover, the tri-holomorphic Killing vector preserves $f$ and
$\omega$.
Hence, these solutions fall into the classification of Gibbons-Hawking
solutions given in Section 3.7. It remains to examine the constraints
imposed on the harmonic functions $H$, $K$, $L$ and $M$ by the
integrability conditions.

To proceed, we note that ({\ref{eqn:integxa}}) implies that
\be
\label{eqn:simmpa}
d \big({K \over H} \big) \wedge d \big( {L \over H} \big) =0
\ee
We shall assume that ${K \over H}$ is not constant, as if  ${K \over
H}$ is
constant then $G^+=0$ and the base is flat; we have already considered
these solutions.
Hence ${L \over H} = {\cal{F}} ({K \over H})$ for some function
${\cal{F}}$;
in fact as a consequence of the harmonicity of $L$, $H$ and $K$
we have $L= \beta H + \gamma K$ for constants $\beta$, $\gamma$.
In addition, it is clear that as $K$ is defined only up to a
shift of a multiple of $H$, we may without loss of generality set
$\gamma=0$,
and so $L = \beta H$. To continue, we note that ({\ref{eqn:integxb}})
implies that
\be
\label{eqn:simmpb}
d \big( {K \over H} \big) \wedge d \big(M+{\beta \over 2}K \big) =0
\ee
and hence $M+{\beta \over 2}K = {\cal{H}} \big( {K \over H} \big)$
for some function ${\cal{H}}$ to be determined. However
({\ref{eqn:integxb}})
also forces ${\cal{H}}$ to be constant, and so without loss of
generality
we obtain $M= -{\beta \over 2}K$. The remaining components of
({\ref{eqn:integxb}}) together with ({\ref{eqn:curvint}}) imply the
following
constraints of on  $H$ and $K$:

\bea
\label{eqn:diffeqnsbiii}
2\rho \delta_{ij} &=&  \nabla_i \nabla_j \big[HK (\beta H^2+ K^2)^{-2}
\big]
\nn
2 \chi \delta_{ij}
&=& \nabla_i \nabla_j \big[ (K^2- \beta H^2) (\beta H^2+K^2)^{-2}
\big]\ ,
\eea
where $\nabla$ denotes the covariant derivative with respect to the
flat metric on $\bR^3$, and
\bea
\label{eqn:cconsta}
\rho &\equiv&- (\beta H^2+ K^2)^{-4} \big[
2HK(\beta H^2 - K^2) (|\nabla K|^2- \beta |\nabla H|^2)
+(K^4-6 \beta H^2K^2 +\beta^2 H^4) \nabla H . \nabla K \big]
\nn
\chi &\equiv& (\beta H^2+ K^2)^{-4} \big[
(K^4-6 \beta H^2 K^2 + \beta^2 H^4) (|\nabla K|^2- \beta |\nabla H|^2)
+8 \beta HK (K^2- \beta H^2)\nabla H . \nabla K \big] \nn
\eea
Note that ({\ref{eqn:diffeqnsbiii}}) imply that $\rho$ and $\chi$ are
constant,
and it is straightforward to show that $\rho=0$ iff $K=0$.
Given these constraints, all of the remaining integrability
conditions then hold automatically.

So, setting $Y_1 =  \rho r^2 + \lambda_i x^i + \sigma$
and $Y_2 =  \chi r^2 + \mu_i x^i + \gamma$ for $\lambda_i$, $\sigma$,
$\mu_i$, $\gamma$
constants, ({\ref{eqn:diffeqnsbiii}}) imply
\bea
\label{eqn:diffeqnsbiv}
HK  (\beta H^2+ K^2)^{-2} &=& Y_1
\nn
(K^2- \beta H^2) (\beta H^2+K^2)^{-2} &=& Y_2
\eea
Hence ({\ref{eqn:diffeqnsbiv}}) fixes $H$ and $K$ according to
\be
\label{eqn:tentata}
K= \delta H
\ee
where $\delta$ satisfies
\be
\label{eqn:deltquad}
\delta^2 -{Y_2 \over Y_1} \delta - \beta =0
\ee
and $H$ is given by
\be
\label{eqn:Hquad}
H^2 = {\delta \over Y_1} (\beta + \delta^2)^{-2}= {Y_1 \over \delta (4
\beta Y_1^2+Y_2^2)} \ .
\ee
With these constraints the five-dimensional spacetime
geometry is simplified. In particular,
\bea
\label{eqn:simpler5d}
f &=& {H \over \beta H^2+K^2}
\nn
\omega_5 &=& {K \over H^2}(\beta H^2+K^2)
\nn
\nabla \cross \bomega &=&2 \beta (K \nabla H - H \nabla K) \ .
\eea
Using these identities, the five-dimensional metric can be written as
\bea
\label{eqn:betnoto}
ds^2 &=& - \beta \big[dx^5- \beta^{-1}K(\beta H^2+K^2)^{-1}dt
+\big(\chi_i-  \beta^{-1}K(\beta H^2+K^2)^{-1} \omega_i\big)dx^i \big]^2
\nn
&+& \beta^{-1} (\beta H^2+K^2)^{-1} \big(dt+\omega_i dx^i\big)^2-(\beta
H^2+K^2)d{\bf{x}}^2
\eea
for $\beta \neq 0$ and
\be
\label{eqn:geteqo}
ds^2=H^2 K^{-4}dt^2+2K^{-1}dt (dx^5+\chi_i dx^i)-K^2 d{\bf{x}}^2
\ee
for $\beta=0$.

\subsubsection{Classifying the Solutions}

We shall neglect cases in which $Y_2$ or $Y_1$
vanish, or for which $Y_2 \propto Y_1$ as this corresponds
to setting $G^+=0$, which we have already classified.
To proceed we shall consider the cases $\beta =0$
and $\beta \neq 0$ separately; in the following $(r, \theta , \phi)$
are standard spherical polar co-ordinates on $\bR^3$.

If $\beta =0$
then from ({\ref{eqn:diffeqnsbiv}}) it is clear that there are two
possibilities. In the first,
\be
\label{eqn:betazero}
K = m \ , \qquad
H = n_i x^i
\ee
for $m$, $n_i$ constants; $m \neq 0$ and $n_i$ not all vanishing.
By changing co-ordinates according as $x^i = m^{-1} {\hat{x}}^i$,
$t= m^3 {\hat{t}}$ and $x^5 = m^{-2} {\hat{x}}^5$ we can without
loss of generality set $m=1$; hence it is clear that this solution
is the maximally supersymmetric plane wave.

Alternatively, one has
\be
\label{eqn:betazerob}
K = {m \over r} \ , \qquad
H = {k \over r} +{n_i x^i \over r^3}
\ee
for $m$, $k$, $n_i$ constants, $m \neq 0$. The five dimensional
metric can be written as
\bea
\label{eqn:adsiiimet}
ds^2 &=& m^{-4} r^4 \big({k \over r}+{n \cos \theta \over r^2} \big)^2
dt^2
+2 m^{-1} r dt \big(dx^5 +(k \cos \theta -{n \sin^2 \theta \over r})
d \phi \big)
\nn
&-&m^2 r^{-2} \big(dr^2 +r^2 (d \theta^2+ \sin^2 \theta d \phi^2) \big)\ .
\eea
It is then convenient to change co-ordinates as $\phi = \phi'- {1 \over
m^3}t'$,
$r=n r'$, $t=n^{-1}t'$. In these new co-ordinates the metric is
\be
\label{eqn:adsiiimetb}
ds^2 = m^{-4} (1+k^2 r'^2)dt'^2 +2m^{-1}r' dt' (dx^5+k \cos \theta d
\phi')
-m^2 r'^{-2}dr'^2 -m^2 (d \theta^2+ \sin^2 \theta d \phi'^2) \ .
\ee
If $k=0$ then this metric is $AdS_3 \times S^2$. If $k \neq 0$ then by
a re-scaling of $r'$ and $x^5$ we may without loss of generality set
$k=1$,
and the curvature invariants of this metric are unchanged from the case
when $k=0$.


Next consider the cases when $\beta \neq 0$. Then $H^2$ and $K^2$
can be written as
\bea
\label{eqn:hiikii}
H^2 &=& -{1 \over 2 \beta} \big[{Y_2 \over 4 \beta Y_1^2+Y_2^2}
\mp {1 \over \sqrt{4 \beta Y_1^2+Y_2^2}} \big]
\nn
K^2 &=& {1 \over 2}  \big[{Y_2 \over 4 \beta Y_1^2+Y_2^2}
\pm {1 \over \sqrt{4 \beta Y_1^2+Y_2^2}} \big]\ .
\eea
It is useful to define $P_{\pm} = \sqrt{Y_2 \pm 2 \sqrt{- \beta} Y_1}$.
Then
\be
\label{eqn:usefulsqident}
{(P_+ \pm P_-)^2 \over P_+^2 P_-^2}=2 \big[{Y_2 \over 4 \beta
Y_1^2+Y_2^2}
\pm {1 \over \sqrt{4 \beta Y_1^2+Y_2^2}} \big]
\ee

Hence, if $\beta <0$ then $P_\pm$ are real, and it follows that
\bea
\label{eqn:betanega}
H &=& {1 \over 2 \sqrt{- \beta}} \big[{1 \over P_-} \mp {1 \over P_+}
\big]
\nn
K &=& {1 \over 2} \big[{1 \over P_-} \pm {1 \over P_+} \big]\ ,
\eea
so ${1 \over P_\pm}$ must be harmonic. This then implies that
there are two sub-cases.
In the first
\be
\label{eqn:ssimpkh}
H ={1 \over \sqrt{- \beta}} \big[ m +{n \over r} \big]
\ , \qquad
K = \big[m - {n \over r}] \ .
\ee
If there exists a point at which $H>0$ and $f>0$
then we require $\beta mn >0$. So $m$ and $n$ have
opposite sign, and the Taub-NUT base space has negative mass.

For this solution,
\be
\label{eqn:auxgodel}
\chi_i dx^i = {n \over \sqrt{-\beta}} \cos \theta d \phi
\ , \qquad  \omega_i dx^i =-4mn \sqrt{- \beta} \cos \theta d \phi
\ee
and hence the five-dimensional metric is
\bea
\label{eqn:ffivdmet}
ds^2 &=& - \beta \big[dx^5+{1 \over 4 m n \beta}(mr-n)dt+{m \over
\sqrt{-\beta}} \cos \theta d \phi \big]^2-{r \over 4 m n \beta}
\big[dt-4mn \sqrt{-\beta}
\cos \theta d \phi\big]^2
\nn
&+&{4mn \over r} \big[dr^2+r^2(d \theta^2+\sin^2 \theta
d \phi^2)\big]\ .
\eea
By changing co-ordinates as $x^5={1 \over \sqrt{-\beta}}(t'-n \psi)$,
$t=4mn \sqrt{- \beta} \psi$ and $r=-{\rho^2 \over 16mn}$,
({\ref{eqn:ffivdmet}}) can be simplified as
\be
\label{eqn:ffivdmets}
ds^2 = \big[d t'+{1 \over 4n}(d\psi- \cos \theta d \phi) \big]^2
-\big[d \rho^2+{\rho^2 \over 4}\big(d \theta^2+\sin \theta^2 d \phi^2
+(d\psi- \cos \theta d \phi)^2\big) \big]
\ee
This is the G\"odel solution.

In the second sub-case $H$ and $K$ have two poles given by
$R_{\pm}=\sqrt{r^2\pm 2 \lambda r \cos \theta + \lambda^2}$
\be
\label{eqn:ssimpkhb}
H ={1 \over \sqrt{- \beta}} \big[ {m \over R_+} +{n \over R_-} \big] \
, \qquad
K = {m \over R_+} -{n \over R_-}
\ee
for $\lambda >0$ constant.
Again, if there exists a point at which $H>0$ and $f>0$ then
$\beta mn >0$, so $m$ and $n$ have opposite sign. For this solution,
it is most convenient to make the following changes of co-ordinate:
\bea
\label{eqn:coordcha}
x^1 &=& \lambda \sqrt{R^2-1} \sin \theta' \cos \phi'
\ , \quad
x^2 = \lambda \sqrt{R^2-1} \sin \theta' \sin \phi'
\ , \quad
x^3 = \lambda R \cos \theta'
\nn
x^5 &=&{1 \over \sqrt{- \beta}} t'
\ , \quad
t ={\sqrt{- \beta} \over \lambda} \psi
\eea
and we obtain
\bea
\label{eqn:solutasum}
H &=& {1 \over \lambda \sqrt{-{\beta}} (R^2- \cos^2
\theta')}((m+n)R+(n-m) \cos \theta')
\nn
K &=& {1 \over \lambda (R^2- \cos^2 \theta')}((m-n)R-(m+n) \cos \theta')
\nn
\chi_i dx^i &=& {1 \over \sqrt{-\beta}(R^2- \cos^2 \theta')}((m+n) \cos
\theta' R^2
+ (m-n) \sin^2 \theta' R - (m+n) \cos \theta') d \phi'
\nn
\omega_i dx^i &=& -{4 m n \sqrt{- \beta} \over \lambda} {(R^2-1) \over
(R^2- \cos^2 \theta')}
d \phi'
\eea
and defining $\chi$ by $\chi = \phi' -{\psi \over 4mn}$ we obtain the
metric
\bea
\label{reqn:orbmpv}
ds^2 &=& (d t' + {1 \over 4}(m^{-1}+n^{-1}) \cos \theta' d\psi +(m-n)R
d \chi)^2
+4mn \big( {dR^2 \over R^2-1}+(d \theta')^2 \big)
\nn
&+&{1 \over 4mn} \sin^2 \theta' (d\psi)^2 +4mn (R^2-1)d \chi^2 \ .
\eea

Next consider the case when $\beta >0$. Then $P_{\pm}$ are complex and
\bea
\label{eqn:betaposa}
H &=& {1 \over 2 \sqrt{- \beta}} \big[{1 \over P_-} - {1 \over P_+}
\big] = {1 \over \sqrt \beta} {\rm Im \big( {1 \over P_-})}
\nn
K &=& {1 \over 2} \big[{1 \over P_-} + {1 \over P_+} \big] =  {\rm Re
\big( {1 \over P_-})} \ .
\eea
Write $P_- = \sqrt{ \tau r^2 + \Omega_i x^i+ \nu}$ where $\tau$,
$\Omega_i$ and $\nu$ are generically complex constants. Requiring
that ${1 \over P_-}$ be harmonic imposes the constraint $\tau \nu
-{1 \over 4}(\Omega_1^2+ \Omega_2^2+ \Omega_3^2)=0$.
There are again two sub-cases.

  In the first, $\tau \neq 0$
and by making appropriate
{\it real} shifts and rotations we can set
\be
\label{eqn:p1sol}
P_- = \zeta \sqrt{r^2+2i \lambda r \cos \theta - \lambda^2}
\ee
for $\zeta \in \bC /  \{0\} $ constant, and $\lambda>0$ a real constant.
Note that if $i \zeta \in \bR$ then the harmonic function
$H$ corresponds to a singular Eguchi-Hanson base space. For this
solution
it is convenient to change coordinates as
\bea
\label{eqn:coordchb}
x^1 &=& \lambda \sqrt{R^2+1} \sin \theta' \cos \phi'
\ , \quad
x^2 = \lambda \sqrt{R^2+1} \sin \theta' \sin \phi'
\ , \quad
x^3 = \lambda R \cos \theta'
\nn
x^5 &=&{1 \over \sqrt{\beta}} \psi
\ , \quad
t ={\sqrt{\beta} \over \lambda} \alpha
\eea
so that, on setting $\zeta = a+ib$ for $a, b \in \bR$, $P_- = \zeta
\lambda (R+i \cos \theta')$ and
\bea
\label{eqn:ingredb}
H&=&-{1 \over \sqrt{\beta}|\zeta|^2 \lambda
(R^2+\cos^2 \theta')}(bR+a \cos \theta')
\ , \quad K={1 \over |\zeta|^2 \lambda
(R^2+\cos^2 \theta')}(aR-b\cos \theta')
\nn
\chi_i dx^i &=& {1 \over \sqrt{\beta} |\zeta|^2 (R^2+ \cos^2
\theta')}(-b \cos \theta' R^2
+a \sin^2 \theta' R-b \cos \theta')d \phi'
\nn
  \omega_i dx^i&=&{\sqrt{\beta} \over \lambda |\zeta|^2}
{(R^2+1) \over (R^2+ \cos^2 \theta')} d \phi'
\eea
and on defining $t'$ by $t' = \phi' + |\zeta|^2 \alpha$ the five
dimensional
geometry is given by
\bea
\label{eqn:urbmpv}
ds^2 &=& -(d \psi+b \cos \theta' d\alpha-{aR \over |\zeta|^2}d t')^2+{1
\over |\zeta|^2}(R^2+1) (dt')^2
\nn
&-& {1 \over |\zeta|^2}\big( {dR^2 \over R^2+1}+(d \theta')^2 \big)-
|\zeta|^2 \sin^2 \theta' (d\alpha)^2 \ .
\eea


In the second sub-case, $\tau =0$ and without loss of generality we can
set
\be
\label{eqn:p2sol}
P_- = \zeta \sqrt{r \sin \theta} e^{i \phi \over 2}
\ee
for
$\zeta \in \bC  /  \{0\}$ constant. By making a rotation, we can
take $\zeta \in \bR / \{0\}$. The metric for this case is given by
taking
\bea
\label{eqn:vbigmet}
H&=& -{1 \over \zeta \sqrt{\beta}} {\sin {\phi \over 2} \over \sqrt{r
\sin \theta}} \ , \qquad
K= {1 \over \zeta \sqrt{r \sin \theta}} \cos {\phi \over 2} \ ,
\qquad \omega_i dx^i = - \zeta^{-2} \sqrt{\beta} r^{-1} \cot \theta dr
\ ,
\nn
\chi_i dx^i &=& - {1 \over \zeta \sqrt{\beta}} r^{1 \over 2} (\sin
\theta)^{-{3 \over 2}} \big(\cos{\phi \over 2}d \theta + \sin \theta
\cos \theta \sin
{\phi \over 2}d \phi \big) \ .
\eea
In fact this solution is once more the maximally supersymmetric
plane wave. To see this first note that by examining
$HK(\beta H^2 +K^2)^{-2}$ it is clear that this solution has $\rho=0$
and hence corresponds to one of the degenerate cases
({\ref{eqn:basea}}),
({\ref{eqn:baseb}}) or ({\ref{eqn:basec}}) for which $K=0$ as discussed
previously. Moreover, the Ricci scalar of the five-dimensional geometry
vanishes, and so the solution must correspond to  ({\ref{eqn:basea}}),
as this is the only case for which the Ricci scalar vanishes.
Hence the five-dimensional geometry is given by
\bea
\label{eqn:ffivedkg}
ds^2 &=& f^2\big( dt+{z \over 6 f^2}({z \over 3}d \phi+2 \sin^{-2}
{\theta \over 2}d \psi) \big)^2
\nn
&-& f^{-1} \big[{9 \over z^2} f df^2 +{z^2 \over 36f}({z \over 3}d \phi
+2 \sin^{-2} {\theta \over 2}d \psi)^2+{f \over 2} \sin^{-4} {\theta
\over 2}
(d \theta^2+ \sin^2 \theta d \psi^2) \big] \ .
\eea
This metric is however equivalent to that given in
({\ref{eqn:nullitya}}) under the co-ordinate transformation
$f={z \over 3}x^1$, $x^2=\sqrt{2} \cot {\theta \over 2} \cos \psi$
,  $x^3=\sqrt{2} \cot {\theta \over 2} \sin \psi$, $t={3 \over
z}{\hat{t}}$
and $x^5={z \over 6} \phi-\psi$ with the identification
$\bomega=(-{3 \over 2},0,0)$ with respect to Cartesian co-ordinates
$x^1$,
$x^2$, $x^3$. Hence the solution is the maximally supersymmetric plane
wave.

\subsection{Summary}

In this section we have determined the most general solutions
preserving maximal supersymmetry. We analysed the solutions
that exist in the null class and the timelike class separately.
In the null class we found flat space, the plane wave and $AdS_3\times
S^2$
and we subsequently saw that each of these also arise
in the timelike class. The base space of the timelike class is
always of Gibbons-Hawking (GH) type. Ignoring flat space,
let us summarise our findings:

$\bullet$ Plane wave: this is in the null class and also in the
time-like class,
where it arises with a smeared Taub-NUT base space (see
\p{eqn:betazero}).
It also arose with the GH base space given in \p{eqn:vbigmet};
it would be interesting
to check whether or not this base space is distinct from smeared
Taub-NUT.

$\bullet$ $AdS_3\times S^2$: this is in the null class and also arises
in the
timelike class with a GH base with a dipole source \p{ghdip}.

$\bullet$ $AdS_2\times S^3$: this has two timelike forms with base space
given by
flat space or singular negative Eguchi Hanson (see discussion following
\p{singeh}).

$\bullet$ Generalised G\"odel: this has two timelike forms with base
space given by flat space or negative mass Taub-NUT
(see the discussion following \p{negtn}).

$\bullet$ Near Horizon BMPV: this has a timelike form with flat base
space.

In addition the timelike analysis revealed three more
geometries with
GH base spaces given in \p{eqn:adsiiimetb} with $k\ne 0$,
\p{reqn:orbmpv} and \p{eqn:urbmpv}. It seems plausible to us that these
are all related to the BMPV solution.
Strictly speaking we analysed necessary conditions for
maximal supersymmetry and
to confirm that these three geometries are indeed maximally
supersymmetric
solutions, one either needs to find a coordinate transformation
confirming
that they are indeed the BMPV solution, or perhaps another
maximally supersymmetric solution listed above, all of which are known
to be explicitly supersymmetric, or alternatively exhibit the
Killing spinors directly.

\sect{$G$-Structures}

We have obtained simple forms for all bosonic solutions
of minimal $D=5$ supergravity that preserve some supersymmetry.
Following \cite{Gauntlett:2001ur,Gauntlett:2002sc},
our method can be related to the notion of $G$-structures.
Recall that a $G$-structure is a reduction
of the principal frame bundle $F$ with structure group $GL(n,\bR)$, to a
subbundle $P$ with structure group $G$ (see eg \cite{joyce}).
Typically such a reduction is equivalent
to the existence of certain globally defined tensors which are invariant
under the group $G$ and it is often
convenient to refer to this set of tensors when
talking about a $G$-structure. In the present setting we assume that
the $D=5$ manifold has a Lorentzian metric $g$ and a spin structure
and hence generically has a $Spin(1,4)$ structure. The existence of a
globally defined Killing spinor $\e$, with isotropy group
$G\subset Spin(1,4)$, gives rise to a $G$-structure.
In particular, various $G$-invariant tensors can be formed from
bilinears in the Killing spinor and these are equivalent
to a $G$-structure.

There are two maximal subgroups of $Spin(1,4)$ that leave a spinor
invariant \cite{bryant}. They are characterised by whether the
corresponding
vector built from the spinor is time-like or null.
In the former case the subgroup is $SU(2)$ while in the latter case
it is $\bR^3$.
In other words for the supersymmetric solutions
admitting a Killing spinor giving rise to a time-like Killing vector
the D=5 geometry admits an $SU(2)$ structure while if it
gives rise to a null killing vector it gives rise to an $\bR^3$
structure. In each case the $G$-structures are characterised by the
algebraic properties satisfied by the metric $g$, the vector $V$
and the two forms $X^i$, which we derived in section 2.

Actually, we should be a little more precise. In the null case, the
vector
is null everywhere and hence the $\bR^3$ structure is indeed globally
defined.
However, in the timelike case the vector can become null, for example
at the horizon of a black hole. Our analysis in section 2
was based on a neighbourhood where $K$ was timelike. In this
topologically
trivial neighbourhood the Killing spinor defines an $SU(2)$
structure. This fact in itself is rather trivial since locally the frame
bundle can always be trivialised. However, the Killing spinor defines
a privileged $SU(2)$ structure satisfying certain differential
conditions
which are not trivial and in fact allow one to deduce the local form of
the solution. The full solution can then be obtained by analytic
continuation.
Note that outside of regions where $K$ becomes null it
defines a global $SU(2)$ structure.

Following \cite{bryant}, these structures can be seen rather explicitly
by exploiting the
isomorphism $Spin(1,4)\simeq Sp(1,1)$. We realize $Spin(4,1)$ as
$2\times 2$
quaternionic matrices $A$ that satisfy $A^\dagger Q A=Q$, where
$$Q=\bigg(\begin{array}{cc}
   1 & 0 \\
   0 & -1
\end{array}\bigg) \ .$$
The spinors are identified with vectors in $\bH^2$ and the action
on spinors is just given by matrix multiplication, $A\cdot s=A s$.
The two types of spinors together with the stabilizer groups are:
\begin{equation}
s=\bigg(\begin{array}{c} 1  \\ 1 \end{array}\bigg)
~~~~\textrm{with stabilizer}~~~ G=\bigg\{\bigg(\begin{array}{cc}
   1+q & -q \\
   q & 1-q
\end{array}\bigg)\bigg|~ q\in Im \bH \bigg\}\simeq \bR{}^3
\end{equation}

and,
\begin{equation}
s=\bigg(\begin{array}{c} r  \\ 0 \end{array}\bigg)
~~~~\textrm{with stabilizer}~~~ G=\bigg\{\bigg(\begin{array}{cc}
   1 & 0 \\
   0 & q
\end{array}\bigg)\bigg|~ q\in Sp(1) \bigg\}\simeq SU(2) \ .
\end{equation}
We can identify $\bR^{4,1}$ with matrices of the form
$m=\bigg(\begin{array}{cc} t & q \\ \bar{q} & t
\end{array}\bigg)$, with $q=x_1+i x_2+j x_3+k x_4 \in \bH$. The
norm of a vector is then given by $det (m)=t^2-q\bar{q}$. Given a
spinor $s=\bigg(\begin{array}{c} p  \\ q \end{array}\bigg)$ we can
construct a vector $V(s)$ by:
\begin{equation}
V(s)=\bigg(\begin{array}{cc} \frac{p\bar{p}+q\bar{q}}{2} & p\bar{q} \\
q\bar{p} & \frac{p\bar{p}+q\bar{q}}{2}
\end{array}\bigg) \ .\nn
\end{equation}
Using this explicit construction we see that the
spinors with stabilizer $SU(2)$ give timelike vectors, while the
ones with $\bR^3$ give null vectors.

Lets consider first the
timelike spinors which define an $SU(2)$-structure.
Let $g_0=dt^2-\sum_{i=1} ^4 dx_i^2$ be the standard Minkowski metric on
$\bR^{4,1}$ and let, \bea V_0&=&\partial_t \nn
X_0^{(1)}&=&dx^{12}-dx^{34}\nn X_0^{(2)}&=&dx^{13}+dx^{24}\nn
X_0^{(3)}&=&dx^{14}-dx^{23}\ . \eea The subgroup of $Spin(1,4)$ that
leaves $(g_0,V_0,X_0^{(i)})$ invariant is $SU(2)_L\subset
SU(2)_L\times SU(2)_R\simeq SO(4)\subset Spin(1,4)$. The three
forms $X^{(i)}_0$ define an almost hyper-K\"ahler structure on the
space transverse to the orbits of the vector $V_0$.
A five dimensional manifold $M$ is said to admit
an $SU(2)$-structure if there exists a non-degenerate metric $g$,
a vector $V$ and three one forms $X^{(i)}$ such that at each
point $p$ there is a map $\alpha : T_p M\rightarrow \bR^{4,1}$
under which $(g,V,X^{(i)})$ are identified with
$(g_0,V_0,X_0^{(i)})$. Using these tensors one can consistently
reduce the structure group to $SU(2)_L$.

\par Let us now discuss the null case. We saw above that we have an
$\bR^3$
structure in this case.
Consider the metric $g_0=2dx^+ dx^- -dx^idx^i$ on $\bR^{4,1}$ and
let: \bea V_0&=&\partial_+\nn X_0^{(i)}&=&dx^-\wedge dx^i \ . \eea
A five dimensional manifold $M$ is said to admit an $\bR^3$ structure if
there exist $(g,V,X^{(i)})$ such that at each point $p$ there
exists a map $\alpha : T_p M\rightarrow \bR^{4,1}$ under which
$(g,V,X^{(i)})$ are identified with $(g_0,V_0,X_0^{(i)})$.
The action of $a^i \in \bR^3$ on
$T^*\bR^{1,4}$  is given by:
\bea dx^-{^{'}}&=&dx^-\nn
dx^+{^{'}}&=&dx^++r^2 dx^-+\sqrt{2}a^i dx^i \nn
dx^i{^{'}}&=&dx^{i}+\sqrt{2}a^i dx^-  \eea
where $r^2=a^i a^i$. Given this explicit action it is clear that
$(g_0,V_0,X_0^{i})$ are left invariant under the action of
$\bR^3$ and thus form an $\bR^3$ structure.

The $G$-structures of interest here can be classified by taking
the covariant derivative of the tensors defining the $G$-structure with
respect to Levi-Civita connection and then decomposing into $G$-modules.
Such a decomposition defines the intrinsic torsion of the $G$-structure.
For example, if all of the $G$-modules vanish, which is equivalent
to the tensors defining the $G$-structure being covariantly constant,
then the Levi-Civita connection has holonomy contained in $G$.
This is what occurs in the D=5 supersymmetric solutions
for vanishing field strength: $\bR^3$ and $SU(2)$ holonomy for the null
and timelike cases, respectively. When the field strength is
non-vanishing
the $\bR^3$ and $SU(2)$ structures are more general and their type
is specified by the differential conditions imposed upon
the tensors that we derived using the Killing spinor equation in
section 2.
For example, the vector $V$ in both cases is not arbitrary but must be
a Killing vector and, for the time-like case, the almost
hyper-K\"ahler structure is actually integrable.

Since we were able to fully characterize the supersymmetric configurations
by the
conditions imposed on the tensors $g,V,X^i$ we conclude that the types
of
$G$-structure that arise in each case provide both necessary and
sufficient
conditions for the existence of supersymmetric configurations of $D=5$
minimal supergravity. This was also true for the class of solutions of
$D=10$ supergravity discussed in
\cite{Gauntlett:2001ur,Gauntlett:2002sc}.

\sect{The G\"odel solution in $D=11$ supergravity}

\label{sec:godel11}

All of the solutions of $N=1$, $D=5$ supergravity can be uplifted on a
flat six
space to obtain solutions of $D=11$ supergravity. Perhaps the most
surprising solution that we found is the maximally supersymmetric
G\"odel solution. Uplifting it to $D=11$ gives another surprise:
naively one would have expected it to still preserve $8$ supersymmetries
but in fact it preserves $20$ supersymmetries.

The solution in $D=11$ can be written
\bea
ds^2&=&-(dt+\omega)^2+ds^2(\bE^{4}) +ds^2(\bE^6)\nn
F&=&-\gamma J\wedge K
\eea
where $J,K$ are K\"ahler forms on $\bE^{4}$, $\bE^6$,
respectively given by
\bea
J&=&dx^1\wedge dx^2+dx^3\wedge dx^4\nn
K&=&dx^5\wedge dx^6+dx^7\wedge dx^8+dx^9\wedge dx^\sharp
\eea
and
\be
\omega={\gamma\over 2}(-x^2 dx^1 +x^1 dx^2 -x^4 dx^3 +x^3 dx^4)
\ee
and hence
\be
d\omega\equiv \gamma J \ .
\ee

It is straightforward to show that this solves the
equations of motion given by
\bea
R_{\mu\nu}-\frac{1}{12}(F_{\mu \s_1\s_2\s_3}F{_{\nu}}{^{\s_1\s_2\s_3}}-
\frac{1}{12}g_{\rho\mu}F^2)&=&0\nn
d*F+\frac{1}{2}F\wedge F&=&0
\eea
where $\epsilon_{0123456789\sharp}=1$.
To determine the amount of supersymmetry we first note that
the conventions we are using have $\Gamma_{0123456789\sharp}=1$
and hence the Killing spinor equation is given by
\begin{equation}\label{killing}
\nabla_\mu\e+\frac{1}{288}[\Gamma{_\mu}{^{\nu_1\nu_2\nu_3\nu_4}}
-
8\delta{_\mu^{\nu_1}}\Gamma^{\nu_2\nu_3\nu_4}]F_{\nu_1\nu_2\nu_3\nu_4}\e
=0 \ .
\end{equation}
Next introduce the obvious orthonormal frame: $(dt+\omega), dx^i, dx^a$
with $i=1,2,3,4$, $a=5,\dots, \sharp$. It is useful to introduce spinors
with three upper indices, $\epsilon^{\cdot\cdot\cdot}$,
each taking the value $\pm$ which specify the chirality with
respect to $\Gamma_{5678}, \Gamma_{789\sharp}$ and $\Gamma_{056}$:
\bea
\Gamma_{5678}\epsilon^{\pm \cdot \cdot}&=&\pm\epsilon^{\pm \cdot \cdot}\nn
\Gamma_{789\sharp}\epsilon^{\cdot \pm \cdot}&=&\pm\epsilon^{\cdot \pm
\cdot}\nn
\Gamma_{056}\epsilon^{\cdot \cdot\pm}&=&\pm\epsilon^{\cdot \cdot\pm} \ .
\eea

We then find that the following constant chiral spinors are
killing spinors:
\be
\epsilon^{+++},\qquad
\epsilon^{+--},\qquad
\epsilon^{-++} \ .
\ee
In addition
\be
\epsilon=\theta^{--+}+(1-\gamma J_{ij}x^i\Gamma^{j56})\theta^{---}
\ee
for constant $\theta^{--+}$,  $\theta^{---}$ are also Killing
spinors. There are no other Killing spinors. Hence
the solution admits precisely $20$ Killing spinors corresponding to
$5/8$ supersymmetry.

The solution has topology $\bR^{11}$ and has closed
time-like curves. Note that one can dimensionally
reduce this solution on the $x^\sharp$ direction
to obtain a type IIA solution and then T-dualise to obtain
a type IIB solution each of which preserves $20$
supersymmetries. These solutions would involve Ramond-Ramond
fields but their simplicity and high degree of symmetry suggests that
it might be interesting to study string propagation in these
backgrounds.

It is interesting to note that G\"odel-like solutions of string theory
have been obtained before. A class of exact string backgrounds with
vanishing
Ramond-Ramond fields was obtained in \cite{horowitz:95}. One of these
solutions (equation 4.11) was interpreted as a rotating
universe. String quantization in this background was studied
in \cite{russo:95,david:02}. Surprisingly, it was not noticed in any of
these papers that the solution contains closed timelike curves or that
it describes a G\"odel-like solution.

\sect{Discussion}

\label{sec:discussion}

We have shown that any supersymmetric solution of minimal
$N=1$, $D=5$ supergravity can be written in one of two simple forms.
Solutions with Killing spinors giving rise to timelike vectors have
$SU(2)$ structures while those giving rise to null Killing vectors
have $\bR^3$-structures. We have also determined the most general
solutions preserving maximal supersymmetry.
We have presented many new solutions but
reasons of space have prevented us from analyzing most of them in any
depth. It is obviously desirable to study our solutions more
carefully to see if they contain any further surprises.

Our work generalizes the analysis
of Tod \cite{Tod} on the minimal $N=2$, $D=4$ theory, which can be
obtained via
dimensional reduction and truncation. The obvious next step is to
undertake
a similar analysis of the minimal $N=1$, $D=6$ supergravity theory,
which also has $8$ supercharges.
In $D=4$, all solutions could be obtained in explicit form but in
$D=5$ this is only possible
in the null case, with the timelike case involving an arbitrary
hyper-K\"ahler manifold. In $D=6$ there is only a null case
\cite{bryant}
and this seems to lead to the supersymmetric solutions
exhibiting $SU(2)\ltimes \bR^4$ structures. It will be interesting to see
if integrable hyper-K\"ahler structures appear or whether something
more general happens.

It would also be interesting
to know the extent to which the method can be extended to non-minimal
theories. Tod examined the minimal $N=4$, $D=4$ theory \cite{tod:95}
(which is
non-minimal in $N=2$ language) but was unable to find all solutions,
or provide (as we have) a simple algorithm for constructing
solutions. However, since minimal $N=1$, $D=5$ reduces to minimal
$N=2$, $D=4$ coupled to a vector multiplet, our results show that the
latter theory must be tractable. Indeed, it is natural to conjecture
that the dimensional reduction of our solutions with Gibbons-Hawking
base space gives the entire timelike class of this $D=4$ theory. The
tractability of this theory suggests examining the general case of
minimal $N=2$, $D=4$ coupled to arbitrarily many vector multiplets.

Gauged supergravities play an important role in the AdS/CFT
correspondence, so a full understanding of the supersymmetric
solutions of these theories is clearly desirable. As far as we know,
no-one has examined this problem even for the simplest case of minimal
$N=2$, $D=4$ gauged supergravity. We expect that this theory and the
minimal $N=1$, $D=5$ gauged supergravity can both be analyzed using the
techniques presented in this paper.


Our interest in understanding the general supersymmetric solutions of a
higher dimensional supergravity theory was motivated in part by a
desire to know whether there exist exotic supersymmetric black holes
in five dimensions. Although we have not found any such solutions, our
general solution is sufficiently complicated that it is not obvious
that such solutions do not exist. It is clearly desirable to have a
uniqueness theorem for supersymmetric black holes in order to justify
the assumptions made in the black hole entropy calculations.\footnote{
We should note that such a theorem has not been proved even for the
simplest
theory admitting supersymmetric black holes, i.e., the minimal $N=2$,
$D=4$ theory. This amounts to proving the conjecture \cite{hartle:72}
that the only black hole solutions in the IWP class are the
Majumdar-Papapetrou multi-black hole solutions, which remains an open
problem \cite{chrusciel:96}. Of course, as soon as a four dimensional
black hole is
made arbitrarily non-extremal, the usual uniqueness theorems apply so
this problem may not be of great physical interest, in contrast with
the higher dimensional case.}
For the minimal $N=1$, $D=5$ theory, proving black hole uniqueness
would involve showing that the only black hole solution belonging to
our general solution is the BMPV solution. A first step might be to
use asymptotic flatness to constrain the base space to be
asymptotically Euclidean and therefore flat, if complete
\cite{gibbons:79}.
However, as we have seen, there is no reason
to suppose that the base space has to be complete and once one permits
incomplete metrics, there are many asymptotically Euclidean
hyper-K\"ahler manifolds.

It is interesting that the maximally supersymmetric G\"odel type
solution
of $D=5$ supergravity
lifts to a solution of $D=11$ supergravity that preserves $20$
supersymmetries.
The simplicity of the solution suggests that there may well be other
similar
solutions preserving exotic fractions of supersymmetry.
Of course the solution does have closed time like curves and thus
its interpretation is not clear. More generally, one of the
conclusions of the work presented here is that closed time-like curves
are
a commonplace amongst supersymmetric solutions. Perhaps there is a good
reason
why such solutions are not relevant in M-theory. On the other hand
maybe they
have a novel dual description waiting to be discovered.

\medskip

\begin{center} {\bf Acknowledgments} \end{center}

We thank Gary Horowitz for drawing Refs \cite{horowitz:95} and
\cite{david:02} to our attention. HSR was supported by PPARC.
JG was supported by EPSRC.

\appendix
\makeatletter
\renewcommand{\theequation}{A.\arabic{equation}}
\@addtoreset{equation}{section} \makeatother

\section{Conventions}

We shall essentially use the conventions of \cite{cremmer:81}, but
it should be noted that unlike that reference our spinors are
commuting spinors, throughout. The metric has signature
$(+,-,-,-,-)$. Tangent space indices will be denoted $\alpha,
\beta \ldots$ and curved indices by $\mu, \nu, \ldots$. The gamma
matrices obey \be
  \{ \gamma_{\alpha},\gamma_{\beta} \} = 2 \eta_{\alpha\beta}
\ee
and satisfy \be
  (\gamma_{\alpha})^{\dagger} = \gamma^{\alpha} = \gamma_0
  \gamma_\alpha \gamma_0.
\ee The antisymmetrization of five gamma matrices is given by \be
\label{eqn:fivegamma}
  \gamma_{\alpha\beta\gamma\delta\epsilon} =
\epsilon_{\alpha\beta\gamma\delta\epsilon}, \ee where
$\epsilon_{01234} = \epsilon^{01234} = +1$.

We use symplectic Majorana spinors $\epsilon_{\alpha}^a$, $a=1,2$,
which are defined as follows. {}First let \be
  \epsilon_a = \epsilon_{ab} \epsilon^b,
\ee where $\epsilon_{ab}$ is antisymmetric with $\epsilon_{12} =
1$. It is convenient to introduce $\epsilon^{ab}$ such that
$\epsilon^{12} = 1$. Now define \be
  \bar{\epsilon}^a = {\epsilon_a}^{\dagger} \gamma_0.
\ee The symplectic Majorana condition is \be
  \bar{\epsilon}^a  = {\epsilon^a}^t C,
\ee where the charge conjugation matrix is real and antisymmetric
and satisfies \be C\gamma_\alpha^tC^{-1}=\gamma_\alpha \ee Note
that \be \label{eqn:interchange}
  \bar{\psi}^a \gamma_{\alpha_1 \ldots \alpha_m} \chi^b = -
  \bar{\chi}^b \gamma_{\alpha_m \ldots \alpha_1} \psi^a,
\ee Given a spinor $\epsilon^a$, one can construct bosonic
quantities \be
  X^{ab}_{\alpha_1 \ldots \alpha_p} \equiv \bar{\epsilon}^a
\gamma_{\alpha_1 \ldots \alpha_p} \epsilon^b. \ee These quantities
obey \be
  X^{ab}_{\alpha_1 \ldots \alpha_p} = -X^{ba}_{\alpha_p \ldots
\alpha_1} \ee as a consequence of equation \p{eqn:interchange}.
{}Furthermore, \be
  \left(X^{ab}_{\alpha_1 \ldots \alpha_p} \right)^* = \epsilon_{ac}
  \epsilon_{bd} X^{cd}_{\alpha_1 \ldots \alpha_p}.
\ee The {}Fierz identity is given by: \be
  \bar{\epsilon}_1 \epsilon_2 \bar{\epsilon}_3 \epsilon_4 = \frac{1}{4}
\left( \bar{\epsilon}_1 \epsilon_4 \bar{\epsilon}_3 \epsilon_2 +
\bar{\epsilon}_1 \gamma_\alpha \epsilon_4 \bar{\epsilon}_3
\gamma^\alpha \epsilon_2 - \frac{1}{2} \bar{\epsilon}_1
\gamma_{\alpha \beta} \epsilon_4 \bar{\epsilon}_3
\gamma^{\alpha\beta} \epsilon_2 \right). \ee Most of the algebraic
identities we recorded in section 2 were obtained by using the
{}Fierz identity with  $\bar{\epsilon}_1 = \bar{\epsilon}^a$,
$\epsilon_2 = \epsilon^d$, $\bar{\epsilon}_3 = \bar{\epsilon}^c$
and then setting in turn $\epsilon_4 = \epsilon^b$, $\gamma_\alpha
\epsilon^b$ and $\gamma_{\alpha \beta} \epsilon^b$. The remaining
identites were obtained using $\bar{\epsilon}_1 =
\bar{\epsilon}^a$, $\epsilon_2 = \gamma_\alpha \epsilon^d$,
$\bar{\epsilon}_3 = \bar{\epsilon}^c$ and $\epsilon_4 =
\gamma_\beta \epsilon^b$.

In various places we parametrise the 3-sphere $SU(2)$ by Euler angles
$(\theta,\phi,\psi)$ with ranges $0\le\theta\le \pi$, $0\le\phi\le
2\pi$ and
$0\le\psi<4\pi$. The detailed parametrisation we use is
described in more detail in Appendix A of \cite{gauntharv}.
The left-invariant or ``right'' one-forms on $SU(2)$ are given by
\bea
\sigma^R_1&=& -\sin\psi d\theta+\cos\psi \sin\theta d\phi\nn
\sigma^R_2&=& \cos\psi d\theta+\sin\psi \sin\theta d\phi\nn
\sigma^R_3&=& d\psi+\cos\theta d\phi
\eea
The right-invariant or ``left'' one-forms are given by
\bea
\sigma^L_1&=& \sin\phi d\theta-\cos\phi \sin\theta d\psi\nn
\sigma^L_2&=& \cos\phi d\theta+\sin\phi \sin\theta d\psi\nn
\sigma^L_3&=& d\phi+\cos\theta d\psi.
\eea
The superscript $R$ ($L$) refers to the fact that the left (right)
invariant
one forms are dual to left (right)
invariant vector fields $\xi^R_i$ ($\xi^L_i$)
which generate right (left) group actions.
We will also refer to $\xi^R_i$
as a right vector field and to $\xi^L_i$ as a left vector field.
The right vector fields are given by
\bea
\xi^R_1&=& -\cot\theta \cos\psi \partial_\psi-\sin\psi \partial_\theta+
{\cos\psi\over \sin\theta}
\partial_\phi\nn
\xi^R_2&=& -\cot\theta \sin\psi \partial_\psi+\cos\psi
\partial_\theta+{\sin\psi\over
\sin\theta}
\partial_\phi\nn
\xi^R_3&=& \partial_\psi
\eea
and the left vector fields by
\bea
\xi^L_1&=& -{\cos\phi\over \sin\theta}
\partial_\psi +\sin\phi \partial_\theta+\cot\theta
\cos\phi\partial_\phi\nn
\xi^L_2&=& {\sin\phi\over \sin\theta}
\partial_\psi+\cos\phi \partial_\theta-\cot\theta \sin\phi
\partial_\phi\nn
\xi^L_3&=& \partial_\phi.
\eea

\section{Integrability condition}
\makeatletter
\renewcommand{\theequation}{B.\arabic{equation}}
\@addtoreset{equation}{section} \makeatother

We record here an integrability condition obtained from the
Killing spinor equation. Taking the second covariant derivative of
the Killing spinor equation and anti-symmetrising we obtain:
\bea
\nabla_{[\rho}\nabla_{\mu]}\e&=& -\frac{1}{4{\sqrt
3}}(\gamma{_{[\mu}}{^{\nu_1\nu_2}}
+4\gamma^{\nu_1}\delta_{[\mu}^{\nu_2})
\nabla_{\rho]}{}F_{\nu_1\nu_2}\e \nn
&+&\frac{1}{48}(\gamma{_{[\mu}}{^{\nu_1\nu_2}}+4\gamma^{\nu_1}
\delta_{[\mu}^{\nu_2})(\gamma{_{\rho]}}{^{\s_1\s_2}}+4\gamma^{\s_1}
\delta_{\rho]}^{\s_2}) {}F_{\nu_1\nu_2}{}F_{\s_1\s_2}\e
\eea
and hence
\bea
\frac{1}{8}R_{\rho\mu\nu_1\nu_2}\gamma^{\nu_1\nu_2}\e
&=&-\frac{1}{4{\sqrt 3}}(\gamma{_{[\mu}}{^{\nu_1\nu_2}}
+4\gamma^{\nu_1}\delta_{[\mu}^{\nu_2})
\nabla_{\rho]}{}F_{\nu_1\nu_2}\e \nn
&+&\frac{1}{48}(-2{}F^2\gamma_{\mu\rho}
-4{}F^2_{\rho\nu}\gamma^\nu{}_\mu +4{}F^2_{\mu\nu}\gamma^\nu{}_\rho
+12{}F_{\mu\nu_1}{}F_{\rho\nu_2}\gamma^{\nu_1\nu_2}\nn
&+&4{}F_{\nu_1\nu_2}{}F_{\nu_3\rho}\gamma_\mu{}^{\nu_1\nu_2\nu_3}
-4{}F_{\nu_1\nu_2}{}F_{\nu_3\mu}\gamma_\rho{}^{\nu_1\nu_2\nu_3}) \e
\eea
where ${}F^2\equiv{}F_{\mu\nu}{}F^{\mu\nu}$ and
${}F^2_{\mu\nu}\equiv{}F_{\mu\s}{}F_{\nu}{}^\s$. Now contracting both
sides of this equation with $\gamma^{\mu}$ and using the Bianchi
identity $R_{\mu[\nu\rho\s]}=0$ we deduce the integrability
condition:
\bea
0&=&(R_{\rho\mu}+2({}F^2_{\rho\mu}-
\frac{1}{6}g_{\rho\mu}{}F^2))\gamma^{\mu}\e\nn
&-&\frac{1}{\sqrt 3}\left[
*(d*{}F+\frac{2}{\sqrt 3}{}F \wedge
{}F)\right] ^{\nu}(2g_{\nu\rho}-\gamma_{\rho\nu})\e\nn
&-&\frac{1}{6{\sqrt
3}}d{}F_{\nu_1\nu_2\nu_3}(\gamma_{\rho}{^{\nu_1\nu_2\nu_3}}
-6\delta_{\rho}^{\nu_1}\gamma^{\nu_2\nu_3})\e
\eea

If we assume that a configuration admits Killing spinors and
satisfies the equation of motion and the Bianchi identity for $F$
we conclude that \be
E_{\mu\nu}\gamma^\nu\epsilon=0 \ee where $E_{\mu\nu}=0$ is
equivalent to the Einstein equations. If we hit this with
$\bar\epsilon$ we deduce that \be\label{intcondone}
E_{\mu\nu}V^\nu=0 \ee On the other hand if we hit it with
$E_{\mu\sigma}\gamma^\sigma$ we conclude that
\be\label{intcondtwo} E_{\mu\nu}E_\mu{}^\nu=0\qquad {\rm no\quad
sum\quad on}\quad \mu \ee

\end{document}